\documentclass[a4paper,12pt]{article} 
\usepackage{graphicx} 
\usepackage{graphics} 
\usepackage{amsmath} 
\usepackage{amsfonts}  
\usepackage{amssymb} 

\newcommand{\dfr}{\widehat{d}} 

\newcommand{\Vsc}{\mathbb{V}} 
\newcommand{\Ds}{{\cal D}} 
\newcommand{\Dv}{{\cal D}_v} 

\def\l{\left}
\def\r{\right}
\def\beq{\begin{equation}}
\def\eeq{\end{equation}}
\def\d{\partial}

\begin{document}


\title{Turbulence and Scale Relativity}

\author{Laurent Nottale and Thierry Lehner \\{\small LUTH, Observatoire de Paris-Meudon,} \\{\small F-92195 Meudon Cedex, France}\\
{\small laurent.nottale@obspm.fr, thierry.lehner@obspm.fr}}
\maketitle

\begin{abstract}
We develop a new formalism for the study of turbulence using the scale relativity framework (applied in $v$-space according to de Montera's proposal). We first review some of the various ingredients which are at the heart of the scale relativity approach (scale dependence and fractality, chaotic paths, irreversibility) and recall that they indeed characterize fully developped turbulent flows. Then we show that, in this framework, the time derivative of the Navier-Stokes equation can be transformed into a macroscopic Schr\"odinger-like equation. The local velocity PDF is given by the squared modulus of a solution of this equation. This implies the presence of null minima $P_v(v_i)\approx 0$ in this PDF. We also predict a new acceleration component in Lagrangian representation, $A_q=\pm \Dv \: \d_v \ln P_v$, which is therefore expected to diverge in these minima. Then we check these theoretical predictions by data analysis of available turbulence experiments:  (1) Empty zones are in effect detected in observed Lagrangian velocity PDFs. (2) We give a direct proof of the existence of the new acceleration component by directly identifying it in the data of a laboratory turbulence experiment. (3) It precisely accounts for the bursts and calm periods of the intermittent acceleration observed in experiments. (4) Moreover, the shape of the acceleration PDF can be analytically predicted from $A_q$, and this theoretical PDF precisely fits the experimental data, including the large tails.  (5) Finally, numerical simulations of this new process allow us to recover the observed autocorrelation functions of acceleration magnitude and the exponents of structure functions.
 \end{abstract}

\section{Introduction}

Turbulence is a complex dynamical phenomenon which involves the coupling of many scales together. Its understanding is ``one of the greatest challenges of modern physics" \cite{Frisch1995}. Turbulent fluids can be seen as spatio-temporal chaotic systems involving a lot of coupled degrees of freedom. This has prevented their description in terms of the theory of low dimensionality chaotic dynamical systems. Moreover, they are out of equilibrium systems. This is manifested through the existence of a cascade of energy flux \cite{Richardson} connecting the various scales. This cascade is conveyed through a multiscale organization of eddies through their fragmentation (direct cascade) or fusion (inverse cascade).

Although the underlying (Navier-Stokes) equations are deterministic, turbulent flows are so complex phenomena that most of their components are usually considered to being fully random, and thus described by stochastic tools. However, one of the main goals of the present paper amounts to show that some components of the Lagrangian acceleration are partly deterministic and behave as pseudo-random variables.

The physical consequences of the cascade were first explored by Kolmogorov (K41) \cite{Kolmogorov1941}. He found that that there is an `inertial' range of scales in which the eddies are too large for viscosity to be important, and too small to retain any effect of large-scale inhomogeneities. The Navier-Stokes equations are invariant to scaling transformations in this inertial domain (see e.g. \cite{Frisch1995}), which ranges from the dissipative small scale ($\eta,\tau_\eta$) to the integral large scale ($L,T_L$) of energy input. In that range, fundamental scaling relations have been found by Kolmogorov for velocity increments, $\delta v \sim \delta x^{1/3}$ (Eulerian) and $\delta v \sim \delta t^{1/2}$ (Lagrangian) \cite{LandauMecaFlu} under the hypothesis of an invariant energy transfer $\varepsilon$ between eddies of different scales. We shall see that this universal Lagrangian K41 scaling also plays a leading role in the present work, interpreted as fractality (of fractal dimension 2) in velocity space \cite{Montera2013}.

One of the main unsolved property of fully developed turbulence is intermittency. It manifests itself as an alternance of calm periods and bursts of intense activity for, e.g., accelerations or velocity increments. One of its signatures is the existence of very large tails of the acceleration probability distribution (PDF), which have been experimentally measured up to more than 50 standard deviations \cite{Mordant2004B}. Another signature is the differences experimentally observed for exponents of structure functions with the K41 expectation. Several stochastic models have been designed to account for this intermittency, beginning with Kolmogorov (K62) \cite{Kolmogorov1962}, in particular multifractal random walks \cite{Castaing1990, Frisch1995, Arneodo1997, Bacry2001, Biferale2004}. These models are based on some experimentally observed specific features (such as correlation functions), but are not dynamical solutions of the Navier-Stokes equations. In the present paper, we suggest an alternative solution to the intermittency problem, involving the effective dynamics and thus accounting for its various characteristics.

The theory of scale relativity, on its side, has been constructed for describing explicitly scale dependent (in particular fractal \cite{Mandelbrot1982}) physical phenomena. For this purpose, it introduces scales in an explicit way, both in variables and equations. In its framework, one looks for the form taken by the equation of dynamics in a fractal and nondifferentiable geometry. One finds that it can be integrated under the form of a Schr\"odinger-type equation. 

Various ingredients of standard quantum mechanics were recovered and demonstrated from this approach \cite{Nottale1993,Nottale2007,Nottale2011}: in particular the wave function is just a manifestation of the velocity field of the fluid of geodesics in a fractal space, and the Schr\"odinger equation is an integral of the equation of geodesics, re-expressed in terms of this wave function. 

But it appeared that the theory could also be applied, as an approximation, to chaotic macroscopic systems \cite{Nottale1993, Nottale1997, Nottale2011}. Indeed, a Schr\"odinger-type equation can be obtained from the equations of dynamics under just three conditions \cite{Nottale1993,Nottale2011}: (i) infinite number of potential paths; (ii) fractality of each path (with fractal dimension 2); (iii) local irreversibility under reflexion of the time increment ($\delta t \leftrightarrow -\delta t$). 

This led us to suggest that a fractal medium could simulate, at some level, a fractal space, and that particles moving in such a medium could therefore acquire macroscopic quantum-type properties: hence we wrote as early as 1993 \cite[Chapt.7]{Nottale1993} ``such a [fractal] medium should show very unusual properties, e.g. a quasi-quantum coherent behavior at macroscopic scales".

The open question since that time was therefore to find such a medium in natural systems, or to build it in a laboratory experiment. However, such systems are expected to be fractal only on a limited range of scales. Two additional constraints should be added to the three above conditions for manifesting such a macroscopic Schr\"odinger regime: (iv) large enough range of fractal scales \cite[Chapt.10]{Nottale2011}; (v) Newtonian dynamics. Indeed, in the application of the theory to quantum mechanics, we considered that space-time was fractal and non-differentiable below the de Broglie scale, without any lower limit  and the dynamics is naturally Newtonian. However, when it is applied to a fractal medium instead of a fractal space, a lower scale is expected for its fractality, and the diffusive aspects of the medium may involve a Langevin-type dynamics instead of a Newtonian one. This reduces the number of systems where such a new physics could be implemented.

One of the natural realm to search for such properties at the observational level was therefore astrophysics, which provides one with both fractal systems on large ranges of scales and Newtonian / Einsteinian dynamics, since their formation and evolution are dominated by gravitation. Then a large body of indirect proofs of a macroscopic Schr\"odinger regime has been revealed for many astrophysical systems on many scales, from planetary systems to extragalactic scales \cite{Nottale1993, Nottale1996, Nottale1997, Nottale2000}, \cite[Chapt.13]{Nottale2011}. Other suggestions of possible implementations of such a new physics have been made in biophysics \cite{Auffray2008, Nottale2008, Turner2017} but also solid state physics where it could be involved in the unusual properties of high temperature superconductors \cite{Nottale2013,Nottale2014b,Turner2017}.

However, a direct proof from a laboratory experiment was still lacking. It has been suggested by L. de Montera that such a proof could be found in a fully developed turbulent fluid \cite{Montera2013}, but in velocity-space instead of position-space. Indeed, the five above conditions are fulfilled for such a fluid, but condition (ii) applies for velocity increments instead of space increments, under the Kolmogorov (K41) inertial scaling relation $\delta v^2 \sim \delta t$ which corresponds to fractal dimension 2 in $v$-space. Moreover, it is known that this relation holds between the integral scale $T_L$ at large scales and the dissipative Kolmogorov scale $\tau_\eta$ at small scales, and that their ratio (i.e., the range of fractal scales) is given by $T_L/\tau_\eta \approx R_\lambda/2 C_0$ in terms of the reduced Reynolds number $R_\lambda=\sqrt{15 Re}$. With $C_0\approx 4$ \cite{Mordant2001}, one obtains a scale ratio of 100 for $R_\lambda=800$, while the turbulence is considered to be fully developed beyond $R_\lambda \approx 500$ \cite{Voth2002}. Finally, being described by the Navier-Stokes equations, the dynamics is essentially Newtonian, at least in the inertial range, although it becomes dominated by the diffusion term $\nu \Delta v$ at small dissipative scales, at which the scaling behavior ceases.

Another strong argument for the application of the scale relativity approach to a fully turbulent fluid is that it is directly adapted to the Lagrangian description of such a fluid (and therefore to a comparison with Lagrangian experiments). Indeed, one obtains the Schr\"odinger-type description \cite{Nottale2007} by identifying the wave function with a manifestation of the velocity field of fractal geodesics \cite{Nottale1993}. 

Numerical simulations of fractal geodesics \cite{Hermann1997}, \cite[Chap.10]{Nottale2011} have been performed in the context of standard quantum mechanics. They have allowed to recover the probability densities which are solutions of the Schr\"odinger equation in a direct way, without writing it explicitly.
These simulations anticipate the application of scale relativity to turbulence. Indeed, in Lagrangian turbulence experiments \cite{Mordant2001,Voth2002}, one follows Lagrangian small particles which are considered as valid tracers of the fluid elements \cite{Qureshi2007}. In our framework, the trajectories of these tracers can then be considered as concrete manifestations of the virtual fractal geodesics introduced in the scale relativity approach.

In the present paper, we first compare the characteristics of turbulent fluids to the various principles underlying the construction of the scale relativity theory (Sec.~\ref{srdvt}). We show in Sec.~\ref{esfta} how the various physical and mathematical tools of scale relativity are fully supported by experimental data of turbulent flows.  We  also briefly review the basic mathematical methods by which one constructs the wave function and the geodesics equation, showing that they are well-known and proven methods widely used in stochastic descriptions of turbulence: it is just their special combination which is specific of scale relativity. In Sec.~\ref{sfftme}, the Schr\"odinger form of the equation of motion is derived, first in position space, then in velocity space for application to turbulence according to de Montera's insight \cite{Montera2013}. The following section~\ref{iatp} describes the main implications and theoretical predictions that one can expect from the new approach, in particular those which can be experimentally put to the test: the main one is the prediction of the existence of a new acceleration component $A_q=\pm \Dv \, \d_v \ln P_v$, where $P_v=| \psi_v|^2$ is the local PDF of velocity given by the square of the modulus of a wave function $\psi_v$, which is solution of a Schr\"odinger-like equation. We list in Sec.~\ref{evote} some experimental observations and results that already come in support of these theoretical expectations and we end by a discussion and conclusion in Sec.~\ref{conclusion}.

\section{Scale relativity description versus turbulence}
\label{srdvt}

The application of the theory of scale relativity \cite{Nottale1993,Nottale2011} to turbulence, which has been initially proposed by L. de Montera \cite{Montera2013}, is supported by many elements \cite{Nottale2013, Nottale2014a, Nottale2014b}. Indeed, let us recall the various ingredients of this theory and put them in correspondence with some recognized characteristics of turbulent fluids.

\begin{itemize}

\item {\bf Scale dependence}. The scale relativity theory (SRT) aims at describing systems which are explicitly dependent on scales. It is well known that this is just the case for a fully developed turbulent fluid.

\item {\bf Scale variables}. In SRT, one describes this scale dependence through the introduction of one or several scale variables. For example, a standard time-dependent function $f(t)$ is replaced by a two-variable function $f(t,\delta t)$ depending on time $t$ and time-scale $\delta t$. 

In turbulent fluids, it is well-known that several physical quantities are explicitly scale dependent in the inertial range, i.e., from the Kolmogorov dissipative small scales to the integral large scales where the energy is injected. For example, in the Lagrangian description in terms of fluid particle trajectories, the accelerations measured for small test-particles are explicitly dependent on the time interval $\tau=\delta t$ (see Fig.~\ref{sigmaadetau}) \cite{LaPorta2001,Voth2002, Mordant2001, Mordant2001T}.

\item {\bf Relativity of scales}. The theory of scale relativity relies on the fact that the various scales are not absolute, but only relative. Indeed, only ratios of scales do have a physical meaning, not a scale by itself. This new relativity is therefore expressed in terms of multiplicative groups instead of the usual additive groups of motion relativity. However, the relevant scale variables being actually given by logarithms of scale intervals ratios, e.g. $\ln(\tau / T)$, one recovers standard additive groups in terms of these logarithmic variables.

The study of turbulent fluid just involves a scale description in terms of such variables, for example $\ln(\tau / T_L)$ or $\ln(\tau / \tau_\eta)$ in Lagrangian representation, where $T_L$ is the Lagrangian integral time-scale  and $\tau_\eta$ the Kolmogorov dissipative time-scale. The scale relativity aspect of turbulence is manifested by the need of such reference scales in the definition of the scale variables, since the dimensioned scale interval $\tau=\delta t$ has no meaning in itself, but only the ratio between this scale and the reference scale.

\item {\bf Chaotic trajectories}. The application of the scale relativity theory to the macroscopic realm is specific of chaotic systems, at time-scales larger than their horizon of predictibility \cite[Chap. 7.2]{Nottale1993}. On these timescales (larger than about 10 to 20 Lyapunov times), the strict determinism is lost and one is led to use a stochastic description. This ensures the first condition underlying the scale relativity description, according to which there is an infinity (or at least a  very large number) of possible trajectories whatever the initial conditions.

It is well known that the fluid element trajectories in a turbulent fluid are chaotic \cite{}. They can even be said to be super-chaotic \cite{}, since the predictibility of individual trajectories in velocity space is lost after some Kolmogorov times (see Fig.~\ref{vxvy}).

\item {\bf Scale laws}. One of the three main conditions upon which the scale relativity descrition relies is the fractal dimension $D_f=2$ of trajectories. In position space, it is expressed by the fact that the space increments and the time increments are no longer of the same order, since $\delta x \sim \delta t^{1/2}$. 

In a similar way, fractality with dimension $D_f=2$ in velocity space is expressed by the relation $\delta v \sim \delta t^{1/2}$. This is just the universal Kolmogorov (K41) scaling law in Lagrangian representation \cite{Kolmogorov1941}. Recall that it can be obtained by simple dimensional analysis based on the assumption that the various scale dependences be driven by the mere energy $\varepsilon$ transfered between scales in the turbulent cascade \cite{LandauMecaFlu} (which is also the energy finally dissipated into heat at Kolmogorov viscous scales). In Eulerian representation, one finds $\delta v \sim \delta x^{1/3}$.

\item {\bf Irreversibility}. The third condition on which the obtention of a Schr\"odinger-type equation relies in the theory of scale relativity is local irreversibility. In the new description, velocities are fractal functions, i.e. explicitly scale dependent functions $v(t,\delta t)$. Their derivative (the acceleration) must be defined from two points, the second point being taken after [i.e., from the velocity increment $v(t+\delta t, \delta t)-v(t,\delta t)$] or before the initial point [$v(t, \delta t)-v(t-\delta t,\delta t)$]. There is no {\it a priori} reason for the two increments to be the same, which leads to a fundamental two-valuedness of the acceleration (that we describe in terms of complex numbers).

It is widely known that the trajectories of fluid elements in a turbulent fluid are irreversible \cite{Falkovich2012}. Here this local irreversibility takes a new meaning, when it is accounted for by this doubling of the acceleration vector and combined with the fractality of trajectories in velocity space.

\item {\bf Newtonian regime}. As we have recalled, we need Newtonian dynamics (linking the force to the second derivative of the variable) to obtain a Schr\"odinger form of the motion equation from  fractality and nondifferentiability. A Langevin regime (in which the action of a force is a velocity instead of an acceleration) does not yield this result \cite[Chap. 10]{Nottale2011}.

The basic equations of fluid mechanics are the Navier-Stokes equations which are clearly of Newtonian nature (i.e., they involve the second derivative of the variable), even if they contain a dissipative viscous term. The same is true after jumping to velocity space: the basic variable becomes the velocity vector and the equation of dynamics is just the time derivative of the Navier-Stokes equations.

\item {\bf Range of fractal scales}. The last condition is that the range of scales involving a $D_f=2$ fractal-type behavior be large enough for the relation $\delta v \sim \delta t^{1/2}$ be fulfilled, at least in an effective way \cite[Sec. 10.3.2]{Nottale2011}.

In turbulent fluids, this means establishing the conditions under which the K41 scaling can be observed. The range of scales where it manifestes is the inertial range, which is limited by the Kolmogorov dissipative scale $\tau_\eta$ and the integral scale $T_L$. Their ratio is given, up to a numerical constant, by the reduced Reynolds number $R_\lambda$:
\beq
\frac{T_L}{\tau_\eta} \approx \frac{R_\lambda}{2 \, C_0},
\eeq
i.e., according to the estimated values of $C_0=4$ to $7$ \cite{Mordant2001, Lien2002, Mordant2004, Sawford2001, Ouellette2006, Sawford2011}, the range of scales is $\approx R_\lambda/10$.  The transition to fully developed turbulence is estimated to be at $R_\lambda \approx 500$, yielding a scale ratio 50, while $R_\lambda \approx 1000$ yields ${T_L}/{\tau_\eta} \approx 100$, as experimentally observed \cite{Mordant2001T, Voth2002}. As we shall see, a well defined (effective) Kolmogorov regime is observed under these conditions (see Fig.~\ref{dv[tau]}). 

 \end{itemize}
 
\section{Experimental support for the application of SRT to turbulence}
\label{esfta}

\subsection{Infinite number of virtual trajectories}
\begin{figure}[!ht]
\begin{center}
\includegraphics[width=16cm]{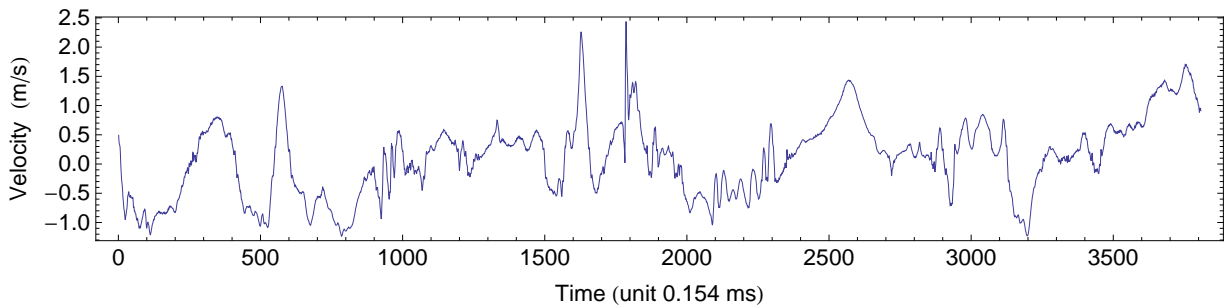} 
\caption{\small{Example of evolution in function of time of the velocity of a Lagrangian particle (Seg3398 of Mordant's experiment man290501).}}
\label{vit3398}
\end{center}
\end{figure}

\begin{figure}[!ht]
\begin{center}
\includegraphics[width=16cm]{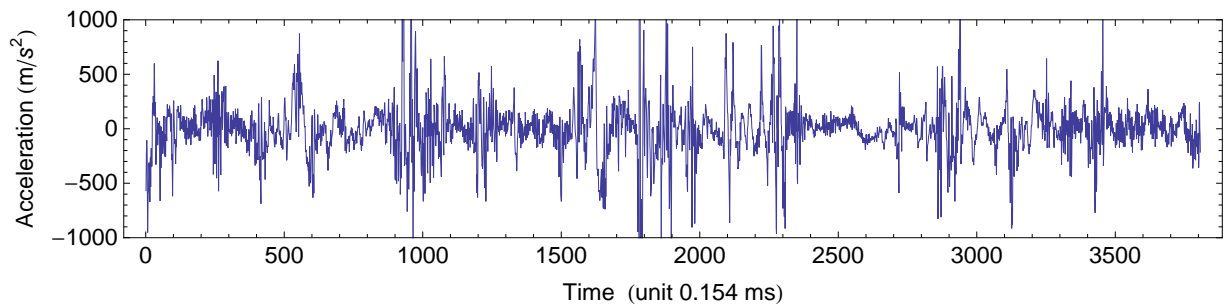} 
\caption{\small{Time evolution of acceleration for Seg3398 of Mordant's experiment man290501. The intermittency is clearly seen in terms of an alternance of quiet periods followed by bursts of fluctuating very high accelerations.}}
\label{acc3398}
\end{center}
\end{figure}

In a turbulent fluid, the trajectories in $v$-space are no longer deterministic. Namely, under the same initial conditions of velocity and acceleration $(v_0,\, a_0)$ [and more generally $(x_0,\,v_0,\, a_0)$], the subsequent evolution of a fluid particle is not determined on time-scales larger than the Kolmogorov dissipative time-scale $\tau_\eta$. This is supported by Lagrangian-type experiments where one follows small particles considered to be valid tracers of the fluid particles (see Figs.~\ref{vit3398} and \ref{acc3398}). In von Karman contra-rotative experiments, its has been shown that particles of size $<\approx100\: \mu$m achieved such valid tracers \cite{Voth2002}. 

\begin{figure}[!ht]
\begin{center}
\includegraphics[width=10cm]{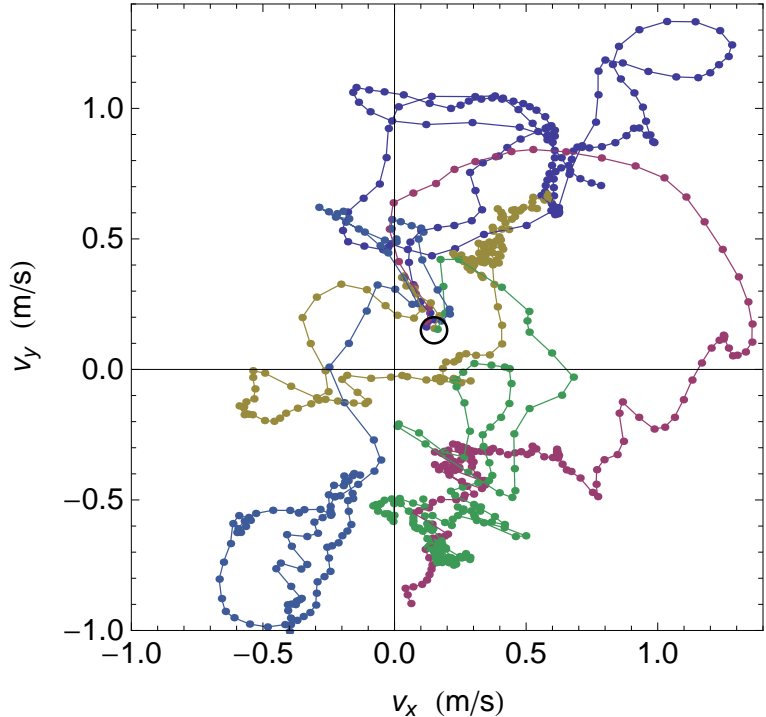} 
\caption{\small{Example of five 2D trajectories in $v$-space (from Mordant's\cite{Mordant} 2D  experiment) starting from close initial conditions (circle) of velocities ($v_x=v_y=0.15$ m/s) and accelerations ($a_x=a_y=300$ m/s$^2$). Each points are separated by a time interval $\tau_u=1/6500$ s $\approx 0.7\, \tau_\eta$. The trajectories are followed on a total time $146 \, \tau_u= T_L$.}}
\label{vxvy}
\end{center}
\end{figure}

We have given in Fig.~\ref{vxvy} an example of five trajectories in $v$-space starting from nearby initial conditions in velocities and accelerations. It is clear on the figure that during the first instants (of some $\tau_\eta$'s), there is a memory of the initial conditions and a partial determinism, after which the trajectories diffuse in a Brownian-like chaotic way. This is in agreement with the observed correlation time of acceleration of $\approx 2.5 \tau_\eta$ \cite{Mordant2001T}.

This supports a description in terms of stochastic scale-dependent variables, $v=v(t,\delta t)$.

\subsection{Scaling laws}
\begin{figure}[!ht]
\begin{center}
\includegraphics[width=12cm]{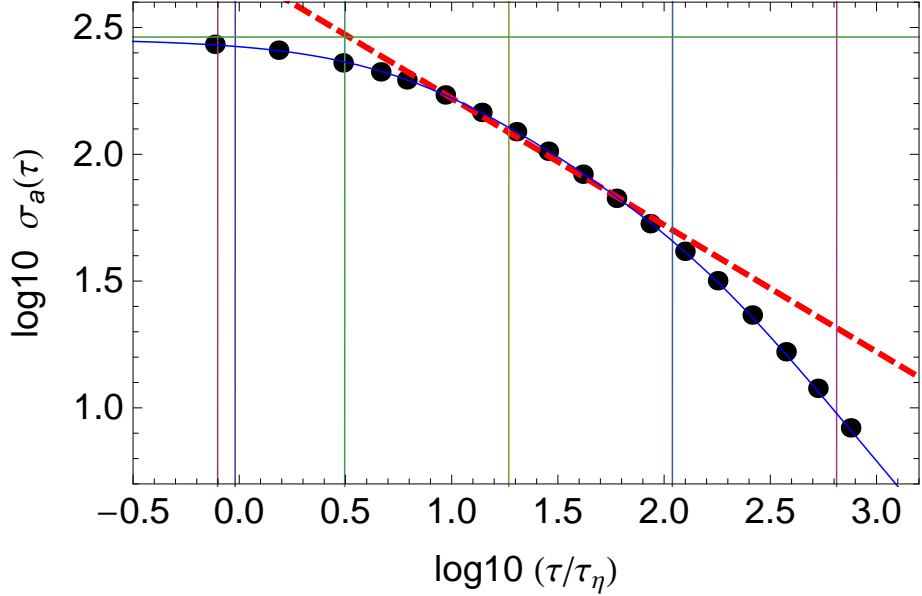} 
\caption{\small{Time-scale dependence of acceleration in Mordant's experiment man290501 (points, $R_\lambda=800$), compared with a K41 scaling law $\sigma_a(\tau) \sim \tau^{-1/2}$ (red dashed line).}}
\label{sigmaadetau}
\end{center}
\end{figure}

\begin{figure}[!ht]
\begin{center}
\includegraphics[width=12cm]{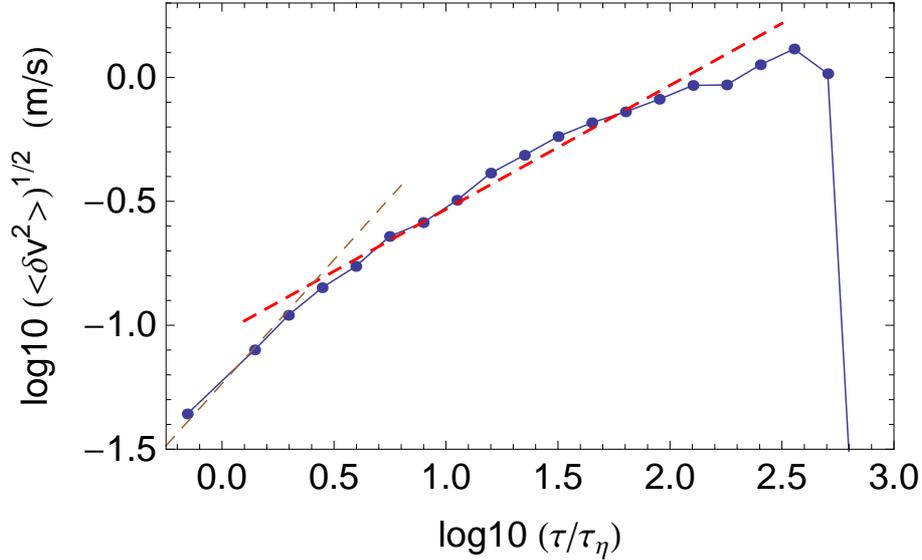} 
\caption{\small{Time-scale dependence of velocity increments in the long segment Seg3398 (3808 velocity values at measurement intervals $\tau_u=1/6500$ s $\approx 0.7\, \tau_\eta$) of Mordant's experiment man290501 (points), compared with the K41 law $\delta v \sim \delta t^{1/2}$ (red dashed line). The small time scale transition is larger than $\tau_\eta$, due to the too large particle diameter $250\, \mu$m. However, the K41 law is achieved in an effective although approximate way on almost two decades.}}
\label{dv[tau]}
\end{center}
\end{figure}

The basic stochastic (and Lagrangian K41) scaling law in the inertial range $\delta v \sim \tau^{1/2}$ (where $\tau$ is the time increment, $\tau=\delta t$) or equivalently for accelerations $\sigma_a(\tau) \sim \tau^{-1/2}$,   can be shown to be present in an effective way in Lagrangian experimental data of fully developed turbulence ($R_\lambda > \approx 500$), see Fig.~\ref{sigmaadetau}.  As can be seen in Fig.~\ref{dv[tau]}, this law is observable locally in individual segments, not only for the full data (3 millions velocity values). This is an important point since, as we shall see, the new structures pointed out here, concerning in particular the PDF of velocities, are purely local.

\begin{figure}[!ht]
\begin{center}
\includegraphics[width=12cm]{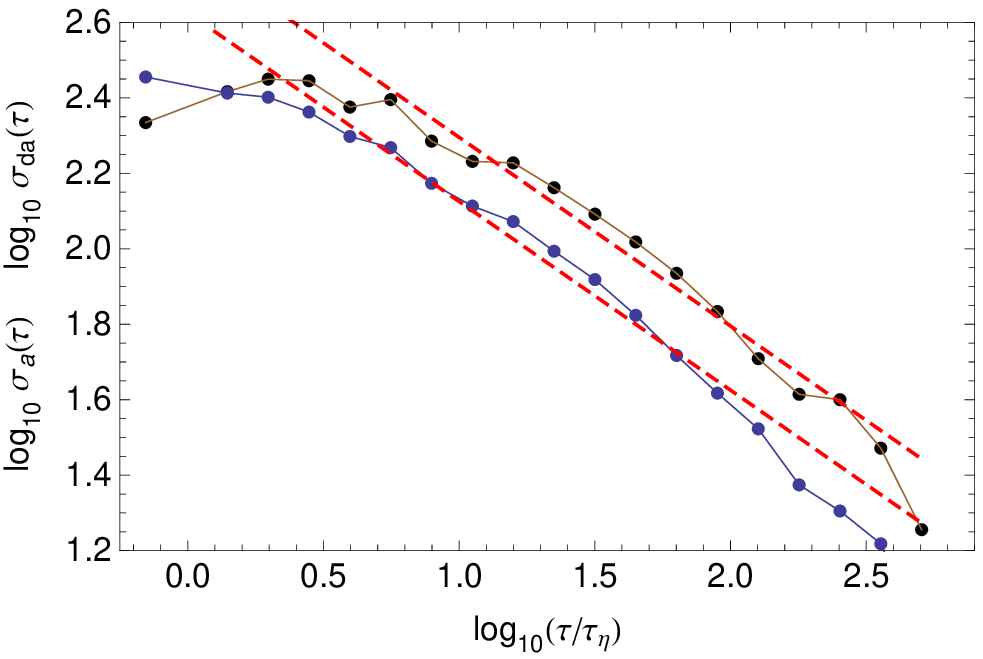} 
\caption{\small{Time-scale dependence of accelerations standard deviations $\sigma_a=<a^2>^{1/2}$ (blue lower points) and acceleration increment standard deviations $\sigma_{da}=<da^2>^{1/2}$ (black upper points) in the long segment Seg3398 (3808 velocity values at measurement intervals $\tau_u=1/6500$ s $\approx 0.7\, \tau_\eta$) of Mordant's experiment man290501. They are  compared with a K41 law $\sim \delta t^{-1/2}$ (red dashed lines). In the inertial range, the amplitude ratio between the two laws is just $\sqrt{2}$, as expected since $a$ is uncorrelated beyond a few $\tau_\eta$'s \cite{Mordant2003,Mordant2004}. }}
\label{ada}
\end{center}
\end{figure}

\begin{figure}[!ht]
\begin{center}
\includegraphics[width=16cm]{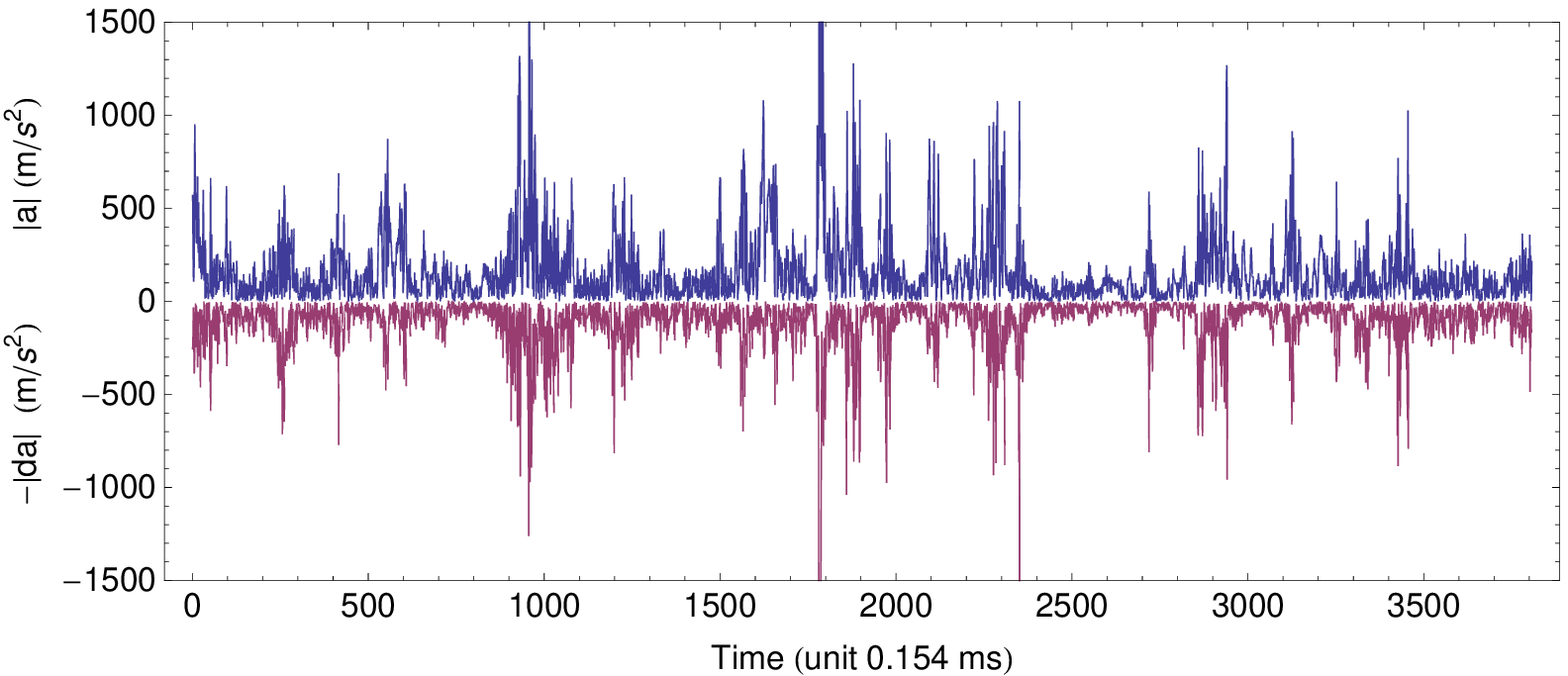} 
\caption{\small{Variation with time of the acceleration $a(t)$ (absolute value, upper part of the figure) compared to its increment $da(t)$ (absolute value, lower part) for Seg3398 (3808 velocity values at measurement intervals $\tau_u=1/6500$ s $\approx 0.7 \; \tau_\eta$) of Mordant's experiment $\# 3$ man290501 \cite{Mordant2001T, Mordant2004}. The acceleration `differential elements' $da$ are of the same order as $a$ itself, in contradiction with the standard differential calculus which assumes $|da| \ll |a|$. }}
\label{ada3398}
\end{center}
\end{figure}

As a direct consequence, the expected scaling law for the acceleration $a$ and for its increment $da$ are $a \sim da \sim \delta t^{-1/2}$. This is also confirmed in the experimental data (see Figs.~\ref{ada} and \ref{ada3398}).

\subsection{Time irreversibility and two-valuedness of acceleration}

We have plotted in Fig.~\ref{ada} a comparison between $a(t)$ and its increments $da(t)$ in function of time for a long  trajectory (Seg3398) in Mordant's experiment man290501 \cite{Mordant2001,Mordant2001T,Mordant2003,Mordant2004}. The time units are $\tau_u=0.7\: \tau_\eta=1/6500$ s. The increments $da$ are here simply measured by finite elements $da= a(t_{i+1})-a(t_i)$ on intervals $t_{i+1}-t_i=\tau_u$.

Two remarkable features appear in this figure. 

(i) While for a standard differentiable curve one would expect $da\ll a$, which is the basis for using the usual differential calculus, it is very clear that this condition is far from being achieved. On the contrary, $da$ is clearly of the same order than $a$ itself, $da \approx a$, and it can therefore not be treated as a standard differential element. 

\begin{figure}[!ht]
\begin{center}
\includegraphics[width=8cm]{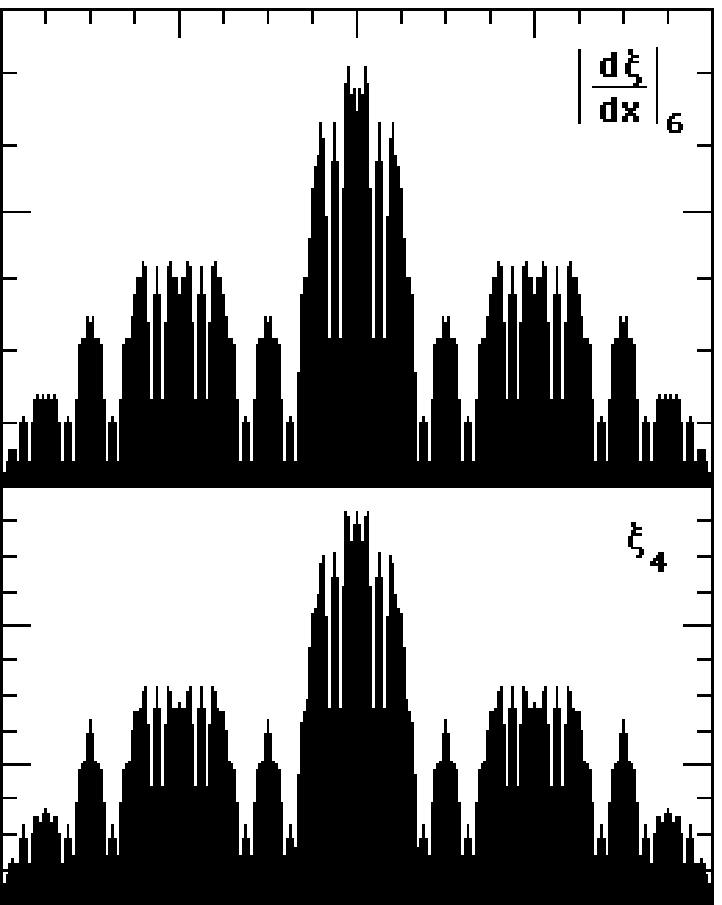} 
\caption{\small{An example, excerpt from Nottale (1993) \cite[Fig.~3.18 p. 79]{Nottale1993} of a fractal function which closely looks like its own derivative. It is constructed from a projection of a fractal curve whose generator is a zig-zag made of 8 segments of length 1/4 (then of fractal dimension $D_f=3/2$). }}
\label{liwos}
\end{center}
\end{figure}

(ii) Moreover, $|a|$ and $|da|$ are not only of the same order, but remarkably similar. Such a possibility has been theoretically anticipated at the beginning of the 90's \cite{Nottale1993}: we had found fractal functions whose derivative were quite similar to the function itself (Fig.~\ref{liwos}). Strictly, such functions $\xi(x,dx)$ are not differentiable and their derivatives are infinite in the limit $dx \to 0$, but by defining them as explicit functions of the increment $dx$, a renormalized derivative $d \xi/dx$ can be defined which is now finite, and which, in the case considered, looks closely like the initial function $\xi$. It appears that the time evolution of the acceleration on a particle trajectory in a fully developped turbulent fluid achieves such a predicted behavior in a laboratory experiment. 

Such a behavior perturbs in an essential way the standard differential calculus. Let us show that it is profoundly linked to irreversibility and that it involves a two-valuedness of the acceleration field.  In the standard non fractal case, one identifies the acceleration $a=dv/dt$ with the first derivative $v'(t)$. This is clear from performing a Taylor expansion:
\beq
\frac{dv}{dt}= \frac{v(t+dt)-v(t)}{dt}=\frac{[v(t)+v'(t) dt + \frac{1}{2} v''(t) dt^2 +...]-v(t)}{dt},
\eeq
so that one obtains
\beq
\frac{dv}{dt}=v'(t)  + \frac{1}{2} v''(t) \, dt + ...= v'(t) +  \frac{1}{2} dv'(t)+ ...
\eeq
For a standard non fractal function, the contribution $ \frac{1}{2}dv'(t) $ and all the following terms of higher order vanish when $dt \to 0$, so that one recovers the usual result $a(t)=dv(t)/dt=v'(t)$. In practice, one does not really take the limit $dt \to 0$, but one considers small time intervals such that $dv'\ll v'$, allowing the same identification (of $a$ and $v'$) in an effective way. 

But in a fully developped turbulent fluid, we have seen that $da \approx a$, so that $dv' \approx v'$ and the term $\frac{1}{2} dv'(t)$ can no longer be neglected with respect to $v'(t)$. Now, the derivative of $v(t)$ involves two points by definition, which may be chosen to be {\it after} (at time $t+dt$, for $dt>0$) or {\it before} the point $v(t)$ (at time $t-dt$).  In the standard differentiable case, the two definitions, lim$_{dt\to 0} [{v(t+dt)-v(t)}]/{dt}$ and  lim$_{dt\to 0} [{v(t)-v(t-dt)}]/{dt}$, coincide. 

But in the non-differentiable case (and for experimental turbulence), the various quantities become explicit functions of the scale interval $dt$, which is no longer considered as tending to 0, but as being variable. The Lagrangian velocity can be described by a fractal function $v(t,dt)$, which an explicit function of two independant variables, time $t$ and the (now non vanishing) differential element $dt$. Then the derivative of velocity can take two forms:
\beq
\frac{d_+}{dt} \: v(t,dt)= \frac{v(t+dt,dt)-v(t,dt)}{dt}, \;\;\;\frac{d_- }{dt}  \: v(t,dt)= \frac{v(t,dt)-v(t-dt,dt)}{dt}.
\eeq
The fact that the increment $dv'(t,dt)$ is no longer negligible with respect to the first derivative $v'(t,dt)$ implies that the two possible expressions for the acceleration are no longer equal:
\beq
a_+=\frac{d_+}{dt} v(t,dt)=v'(t,dt)  + \frac{1}{2} \;dv'(t,dt)  , \;\;\; a_-=\frac{d_-}{dt} v(t,dt)=v'(t,dt)  - \frac{1}{2} \; dv'(t,dt).
\eeq
Therefore, there is a two-valuedness of the possible values of the acceleration, since in general $a_+ \neq a_-$. Note that this two-valuedness does not come from time reversal. The two ($+$) and ($-$) values are {\it not} a forward and a backward acceleration. In both cases time goes, as physically expected, from past to future: one goes from one definition to the other by the transformation ($dt \to -dt$), not ($t \to -t$). Nor is it a left and right derivative of a standard function: this two-valuedness is specific of fractal functions which are explicitly dependent on the finite differential element $dt$, identified with a resolution interval.

\subsection{Methods}
\label{methods}

The methods by which these various conditions are mathematically implemented in the theory of scale relativity are actually standard, widely used, methods, in particular in the domain of turbulence studies (stochastic calculus, Ito formulae, fractals and mutifractals, complex numbers). The new ingredient that leads to original results (namely, a Schr\"odinger form for the derivative of Navier-Stokes equations after integration) is just their combination, which has not been considered up to now in this form in fluid mechanics.

 The infinity of the number of possible trajectories naturally leads to a stochastic description. The use of stochastic differential equations (SDE's) is a standard method in turbulence studies \cite{Sawford1991,Borgas1994,Pope2002} and is also a basis of the present approach.

The Ito calculus is particularly adapted to derivation and integration of stochastic variables when second order differential elements intervene in an explicit way. This is just the case for fluctuations in turbulence characterized in the inertial range by different orders for the velocity and time differential elements, according to Kolmogorov scaling $\delta v^2 \sim \delta t$. It is a standard tool in turbulence studies. 
We also naturally use it to build the ``covariant" total derivative which generalizes the Euler derivative $d/dt=\d/\d t + v. \nabla$ (which is specific of fluid mechanics) to fractal geometry. 

Fractals and multifractals are now standard tools for building models in the study of turbulence. Here we use this concept in a somewhat different way, since we consider the turbulent fluid as a medium which is fractal in velocity-space and plays the role of a fractal space for the particles moving into it \cite{Nottale1993}.

Complex numbers are used in the theory of scale relativity as a natural description of the algebra doubling
imposed by the two-valuedness of accelerations. We combine the (+) and (-) accelerations in terms of a doublet $(a_+,a_-)$, then in terms of complex numbers $ {\cal A}=(a_+ +a_-)/2 - i (a_+ -a_-)/2$.

\section{Schr\"odinger form for the motion equation}
\label{sfftme}

\subsection{General argument}
As we shall recall hereafter, the combination of the three conditions (infinite number of possible trajectories, scaling law of fractal dimension 2 and local irreversibility) leads to give to the energy equation (integral of a Newtonian equation of dynamics) a Schr\"odinger-type form \cite{Nottale1993}. This result is obtained from a general mathematical argument, which does not depend on the nature of the variables. We shall first remind how it is obtained in position space, as made in previous publications whose goal were to describe the dynamics in a fractal space \cite{Nottale2011}. This case corresponds to a fundamental fractality relation $\delta x \sim \delta t ^{1/2}$ which characterizes paths of fractal dimension 2 (as expected for stochastic Markovian paths being neither correlated nor anticorrelated).

However, in the case of laboratory turbulence, as pointed out by L. de Montera \cite{Montera2013}, the fractality is not in position space, but in velocity space, as shown by the Lagrangian K41 scaling relation $\delta v \sim \delta t^{1/2}$. He then made the remarkable suggestion of applying the scale relativity theory to turbulence in $v$-space instead of $x$-space. In  other words, the fundamental space variable becomes the velocity $v$, and the equivalent of the successive derivatives $(x,\:v,\:a)$ become $(v,\: a,\: \dot{a})$. Therefore, the Navier-Stokes equation remains unaffected by this new dynamics, but its derivative takes, after integration, the form of a Schr\"odinger-type equation.

\subsection{Scale relativity theory in position space: a short reminder}
The laws of motion are obtained in the scale relativity theory by writing the fundamental equation of dynamics (which reduces to a geodesic equation in the absence of an exterior field) in a fractal space (more generally, space-time). As we have seen, the non-differentiability and the fractality of coordinates implies at least three consequences \cite{Nottale1993, Nottale2011}:

\noindent (1) The number of possible paths is infinite. The description therefore naturally becomes non-deterministic and probabilistic. These virtual paths are identified with the geodesics of the fractal space. The ensemble of these paths constitutes a fluid of geodesics, which is therefore characterized, as a first step, by a velocity field $V(X,t)$ in Eulerian representation. The individual geodesics, which describe the possible trajectories of particles and are therefore of Lagrangian essence, are linked to this Eulerian field by the relation $dX_\alpha / dt= V[X_\alpha(t),t]$, for each geodesic labelled by $\alpha$.

\noindent (2) Each of these paths is itself fractal with dimension 2. The velocity field is therefore a fractal function, $V(x,t,dt)$, explicitly dependent on resolutions and divergent when the scale interval tends to zero (this divergence is the manifestation of non-differentiability).

\noindent (3) Moreover, the non-differentiability also implies a two-valuedness of this fractal function, $\{V_+(X,t,dt), \; V_-(X,t,dt)\}$. Indeed, two definitions of the velocity field now exist, which are no longer invariant under a transformation $|dt| \to -|dt|$ in the nondifferentiable case. 

These three properties of motion in a fractal space lead to describing the geodesic velocity field in terms of a complex fractal function,
\beq
{\cal V}=\frac{ V_+ + V_-}{2} - i \; \frac{V_+ - V_-}{2}.
\eeq
The elementary displacements along these geodesics can be described in terms of stochastic differential equations (SDE), decomposing them as the sum of a mean and a fluctuation. They take two forms which read (in Lagrangian representation)
\begin{eqnarray}
dX_+ = v_{+} \; dt+d\xi_{+},\\
dX_- = v_{-} \; dt+d\xi_{-}.
\label{eq.20bis}
\end{eqnarray}
The scale dependence of the fractal fluctuations reads $d\xi_{\pm}=\zeta_{\pm} \, \sqrt{2 \cal{D}}  \, |dt|^{1/2}$ , with
 $\zeta_{\pm}$  a dimensionless stochastic variable such that $<\!\!\zeta_{\pm}\!\!>=0$ and $<\!\!\zeta_{\pm}^2\!\!>=1$. The parameter $\cal{D}$ characterizes their amplitude.

Using Ito calculus, 
these various effects can be combined in terms of a total derivative operator \cite{Nottale1993} which generalizes to fractal geometry the Euler total derivative (and therefore applies to Eulerian representation) :
\begin{equation}
\frac{\widehat{d}}{dt} =  \frac{\partial}{\partial t} + {\cal V}. \nabla -i {\cal  D} \Delta.
\label{dercov}
\end{equation} 

Various equivalent representations of the equations of dynamics become now possible by using this tool.

\paragraph{Geodesic representation}

The first representation, which can be considered as the root representation, is the geodesic one. The two-valuedness of the velocity field is expressed in this case in terms of the complex velocity field $\Vsc=V-i U$. It implements what makes the essence of the principle of relativity: namely, the equation of motion, once written in terms of the covariant derivative Eq.~\ref{dercov}, takes the form of a free inertial equation devoid of any force:
\beq
\frac{\dfr }{dt}{\cal V}=0,
\eeq
where the `covariant' derivative operator ${\dfr}/{dt}$ includes the terms which account for the effects of the geometry of space -- more generally of space-time in the `relativistic' case (see \cite{Nottale2011} and references therein).
In presence of an exterior potential, the equation of dynamics takes Newton's form:
\beq
m \: \frac{\dfr }{dt}{\cal V}=- \nabla \phi.
\label{Newton}
\eeq

\paragraph{Schr\"odinger representation}
A wave function $\psi$ can be introduced \cite{Nottale1993} as a re-expression of the action ${\cal S}=-2 i \Ds \ln \psi$ (which is now complex since the velocity is complex). As a consequence of the fundamental canonical relation ${\cal P}=m {\cal V}= \nabla {\cal S}$, the velocity field of geodesics is related to the wave function according to:
\beq
{\cal V}= -2 i \Ds \, \nabla  \ln \psi.
\eeq
This means that the two-valuedness of the velocity field (coming from non-differentiability) is expressed in terms of two quantities, the squared modulus $P$ and the phase $\theta$ of this wave function. 

By replacing the complex velocity in Eq.~\ref{Newton} by this expression, one finds that the equation of motion can be integrated under the form of a Schr\"odinger equation \cite{Nottale1993},
\beq
 {\cal D}^2 \Delta \psi + i  \, {\cal D}\frac{{\d}}{\d t}  \psi - \frac{\phi}{2 m} \,\psi = 0.
 \label{Schro}
\eeq

\paragraph{Fluid representation with quantum potential}

By decomposing the complex wave function $\psi = \sqrt{P} \times e^{i \theta/2\Ds}$  in terms of a modulus $\sqrt{P}$ and a phase $\theta/2\Ds$ (related to the real part of the velocity field by the relation  $V= \nabla \theta $), one can give this equation the form of hydrodynamics equations including a quantum potential \cite{Nottale1993,Nottale2009, Nottale2011, Nottale2012}.

Indeed, the imaginary part of the Schr\"odinger equation becomes a continuity equation:
\beq
\frac{\d P}{\d t} + \text{div} ( P V)=0.
\eeq
The real part takes the form of an Euler equation, 
\beq
m \l( \frac{\d}{\d t} + V. \nabla\r) V=- \nabla (\phi +Q),
\eeq
where $Q$ is an additional potential energy that depends on the probability density $P$:
\beq
Q=-2 m \Ds^2 \: \frac{ \Delta \sqrt{P}}{\sqrt{P}}.
\eeq
The additional quantum potential is obtained here as a direct manifestation of the fractal geometry of space, in analogy with Newton's potential emerging as a manifestation of the curved geometry of space-time in Einstein's relativistic theory of gravitation. 

\subsection{Velocity space: application to laboratory turbulence}
The application of the scale relativity approach to fluid turbulence in the inertial range (where the relation $\delta v \sim \delta t^{1/2}$ strictly holds) amounts to just shift the variables $(x,\:v,\:a)$ to $(v,\: a,\: \dot{a})$. Now, the velocity $V$ is the primary variable, while
the fundamental local irreversibility leads to a two-valuedness of the acceleration field, $(A_+,\: A_-)$, which is represented in terms of a complex acceleration,
\beq
 {\cal A}=\frac{ A_+ + A_-}{2} - i \; \frac{A_+ - A_-}{2}= A_R -i A_I.
\eeq
In the inertial range, and neglecting for the moment the Langevin term $-v/T_L$, the new Lagrangian description starts with two stochastic differential equations in $v$-space:
\begin{eqnarray}
dV_+ = A_{+} \; dt+dV_{\xi+}\\
dV_- = A_{-} \; dt+dV_{\xi-}
\label{SDE}
\end{eqnarray}
where the scale dependence of the stochastic fluctuation reads, in agreement with the K41 scaling law,
\beq
dV_{\xi\pm}=\zeta_{\pm} \, \sqrt{2 \Dv}  \, |dt|^{1/2}.
\eeq
Such linear SDE are known to yield Gaussian processes (see \cite{Pope2002} and references therein).  Therefore, the reduced variable $\zeta_{\pm}$  is taken here to be a dimensionless Gaussian stochastic variable such that $<\!\!\zeta_{\pm}\!\!>=0$ and $<\!\!\zeta_{\pm}^2\!\!>=1$. The parameter $\Dv$ characterizes the amplitude of the fluctuations. 

Using  Ito calculus, these various effects can be combined  in terms of a total derivative operator acting in $v$-space:
\begin{equation}
\frac{\widehat{d}}{dt} =  \frac{\partial}{\partial t} + {\cal A}. \nabla_v -i  \Dv \Delta_v.
\label{deco}
\end{equation} 

Let us now consider the Navier-Stokes equation in Lagrangian form (reduced to the Euler equation in the inertial range by neglecting for the moment the viscous term), ${d v}/{dt}=F$,
and derivate it with respect to time:
\beq
 \frac{d a}{dt}=\dot{F}.
\eeq
In order to account for the various effects described here, one replaces $d/dt$ by the new total derivative operator ${\widehat{d}}/{dt}$. Assuming that the force in $v$-space contains a contribution that is the $v$-gradient of a potential $\Phi_v$, one gets for this contribution the equation:
\beq
\frac{\dfr }{dt} {\cal A}=-\nabla_v \Phi_v.
\eeq
The existence of such a leading contribution is supported by the description of the cascade in terms of eddies. Indeed, the pulsating \cite{LandauMecaFlu} or oscillatory nature of the motion in eddies for each variable leads us to naturally formalize them as oscillators. This allows us to insert the eddies directly in the equation of motion under the form of oscillator potentials. This can be easily done as well in $x$-space as in $v$-space. As we shall see, this method becomes particularly useful with the Schr\"odinger form of the motion equation, since oscillator solutions (harmonic and anharmonic) are well known and largely studied in quantum mechanics.

We now introduce a wave function $\psi_v$ as a re-expression of the action ${\cal S}_v$ which is now complex (since the dynamical variables are complex):
\beq
\psi_v= e^{i {\cal S}_v/2 \Dv}.
\eeq
It can be decomposed in terms of a modulus and a phase:
\beq
\psi_v=\sqrt{P_v} \times e^{i \theta_v/2 \Dv}.
\eeq
The main point here is that the PDF of velocities is given by the square of the modulus of the wave function,  $P_v(v)= |\psi_v|^2$, while its phase is linked to the real part of the complex acceleration through the relation $A_R=\nabla_v \theta_v$. The constant $2\Dv$ is therefore the macroscopic equivalent in $v$-space of the constant $\hbar$ of standard quantum mechanics (for $m=1$).

Finally, the derivative of the fluid equations takes, after integration, the form of a macroscopic Schr\"odinger equation \cite{Montera2013,Nottale2013,Nottale2014b}:
\beq
\Dv^2 \; \Delta \psi_v + i  \, \Dv \frac{{\d}}{\d t}  \psi_v - \frac{\Phi_v}{2} \,\psi_v = 0.
 \label{Schrov}
\eeq
This Schr\"odinger equation does not contain all the contributions to the dynamics, but it can be considered as a kind of kernel to which the other effects (non potential terms, Langevin term, viscosity, etc.) can be added (see \cite{Nottale2009}).

The complex acceleration field writes, in terms of the wave function,
\beq
{\cal A}=-2i \:\Dv\: \nabla_v \ln \psi_v,
\eeq
so that we are now able to establish the expression for the two new acceleration components $A_+$ and $A_-$:
\begin{eqnarray}
A_+=+ \Dv\; \frac{\d_v P_v}{P_v} + \d_v \theta_v,\\
A_-=- \Dv\; \frac{\d_v P_v}{P_v}  +\d_v \theta_v.
\label{AA}
\end{eqnarray}
In many situations which may be relevant to the turbulence case, in particular for a harmonic oscillator potential expected to describe the largest eddies of the turbulent cascade as a first approximation, the solutions of the Schr\"odinger equation are real \cite{LandauMQ}, i.e. $\theta_v\approx$ cst and then $\d_v \theta_v \approx 0$. Under this approximation (which is supported by the experimental data), the new acceleration can then be written as:
\beq
A_q= \pm \Dv\; \frac{\d_v P_v}{P_v}.
\eeq
The SDE's which describe the possible trajectories of fluid particles then writes
\beq
dV_\pm = -\frac{V}{T_L} \;dt  \pm \Dv\: (\d_v \ln P_v) \; dt+dV_{\xi\pm}, \;\;\;\; <dV_{\xi\pm}^2(\tau)>=2 \Dv \tau,
\label{SDE2}
\eeq
where $\tau=\delta t$ is a varying time-scale interval and $\Dv$ is a constant diffusion coefficient.
We have added a Langevin large scale contribution in order to account for the velocity auto-correlation and to connect our approach to the SDE usually written in stochastic models of turbulence \cite{Sawford1991, Borgas1994, Pope2002}:
\beq
dV= -\frac{V}{T_L} \;dt+dW_{\xi}, \;\;\; <dW_{\xi}^2(\tau)>=2 D_0 \tau.
\eeq

There are some important differences between the two models: 

(i) The new contribution $A_q$ is expected to now carry a part of the variance of the acceleration. As we shall see in what follows, this is supported by experimental data : we find that about $85\%$ of the variance is accounted by $A_q$ alone. Therefore the diffusion coefficient in the new model is expected to be smaller than in the standard one, $\Dv \ll D_0$; 

(ii) There are two SDE's instead of one because of the two-valuedness $\pm$ (but they can be combined in terms of a single complex number SDE);

(iii) If one wants to account for intermittency and highly non-Gaussian large tails of the acceleration PDF from the usual SDE, one should introduce a non-Gaussian probability distribution of the stochastic fluctuation. In this framework, the very origin of this non-Gaussianity is still unknown, even though it can be described by multifractal models \cite{Castaing1990, Frisch1995, Arneodo1997, Bacry2001, Biferale2004} or by making the model coefficients themselves stochastic processes \cite{Beck2001}. In the new scale relativity framework, we shall show that the wide tails of the acceleration PDF can be obtained as an effect of the new acceleration component $A_q$ {\bf alone} (see Sec.~\ref{PDFofA}). The stochastic fluctuation has therefore no reason to remain non-Gaussian, and can be described by a simple Brownian motion in $v$-space. As we shall see in the following, this is supported by experimental data, since we find that the calm periods of the intermittent acceleration, which are just the zones where $A_q \approx 0$, i.e. of pure fluctuations, are perfectly Gaussian (see Sec.~\ref{gaussianity} and Fig.~\ref{Gauss}).

Finally, the full process in the scale relativity approach to turbulence is described by a combination of two coupled equations, for $A_\pm$ a SDE, and for the derivative of $A_\pm$ the Schr\"odinger equation (which is just an expression of the derivative of the Navier-Stokes equation written in fractal $v$-space, ${\dfr } {\cal A}/{dt}=-\nabla_v \Phi_v$, in the inertial domain):
\begin{eqnarray}
A_\pm = -\frac{V}{T_L}   \pm \Dv\: (\d_v \ln P_v) +A_{\xi\pm}, \;\;\;  <\!\!A_{\xi \pm}^2(\tau)\!\!> =2 \Dv/ \tau,\\
\Dv^2 \; \Delta \psi_v + i  \, \Dv \frac{{\d}}{\d t}  \psi_v - \frac{\Phi_v}{2} \,\psi_v = 0, \;\;\;\;  P_v=|\psi_v|^2,
\label{system}
\end{eqnarray}
where $A_{\xi\pm}$ is the time derivative of the stochastic process $dV_{\xi\pm}$. We shall give, in what follows, examples of numerical simulations of solutions to these coupled equations.

This system can be put in correspondence with some standard stochastic approaches, in which, instead of a mere Langevin SDE, two-level equations (an ODE for the acceleration $a$ and an SDE for its derivative $\dot{a}$) are also used \cite{Krasnoff1971,Sawford1991,Pope2002}.

\paragraph{Quantum-classical transition}
\label{Qcltrans}
One of the main features of standard quantum mechanics is the existence of a quantum to classical transition around the de Broglie length-scale. A system becomes quantum for length-scales smaller than $\lambda_{dB}=\hbar/(m <\!\!v^2\!\!>^{1/2})=\hbar/m \sigma_v$ and classical at larger scales. 

The equivalent transition in the case of fluid turbulence is naturally a velocity-scale, $\delta v_{dB}=\hbar_v/<\!\!a^2\!\!>^{1/2}=\hbar_v/\sigma_a=2 \Dv/\sigma_a$, where $\sigma_a=<\!\!a^2\!\!>$ is the standard deviation of acceleration. The role of the Planck constant $\hbar_v$, here defined in $v$-space, is now played by the constant $2 \Dv$ (for $m=1$).

 However, the existence of a lower time-scale, the Kolmogorov dissipative scale  $\tau_\eta$, transforms this velocity interval into an acceleration $a_{dB}=\delta v_{dB}/\tau_\eta$. We know that, under the K41 regime, the acceleration variance is given by \cite{Heisenberg1948,Yaglom1949}:
 \beq
 \sigma_a^2= A_0 \frac{\varepsilon}{\tau_\eta},
 \eeq
where $A_0$ is a numerical constant which has been estimated to be, in fully developped turbulence ($R_\lambda > \approx 500$),  $A_0\approx 4$ in DNS and $A_0\approx 6$ from experimental data \cite{Voth2002}, while it is smaller in Mordant et al data \cite{Mordant2004}, $A_0\approx 1$, probably due to the large particle size \cite{Qureshi2007}.
 We also know that, in the K41 regime,
\beq
2 D_0=C_0 \varepsilon= \frac{2\, \sigma_v^2}{T_L}.
\eeq
Then we find from these various relations that the de Broglie-like acceleration transition is given by:
\beq
a_{dB}=\frac{\Dv}{D_0} \frac{ C_0}{A_0}\;  \sigma_a \approx \sigma_a.
\eeq
Therefore, we expect macroscopic quantum-like effects to manifest at accelerations $a<\approx \sigma_a$, while the large accelerations remain ``classical". As we shall see, experimental data support this expectation (see Figs.~\ref{neufsegs} and \ref{vat}).

\section{Implications and theoretical predictions}
\label{iatp}

\begin{figure}[!ht]
\begin{center}
\includegraphics[width=9cm]{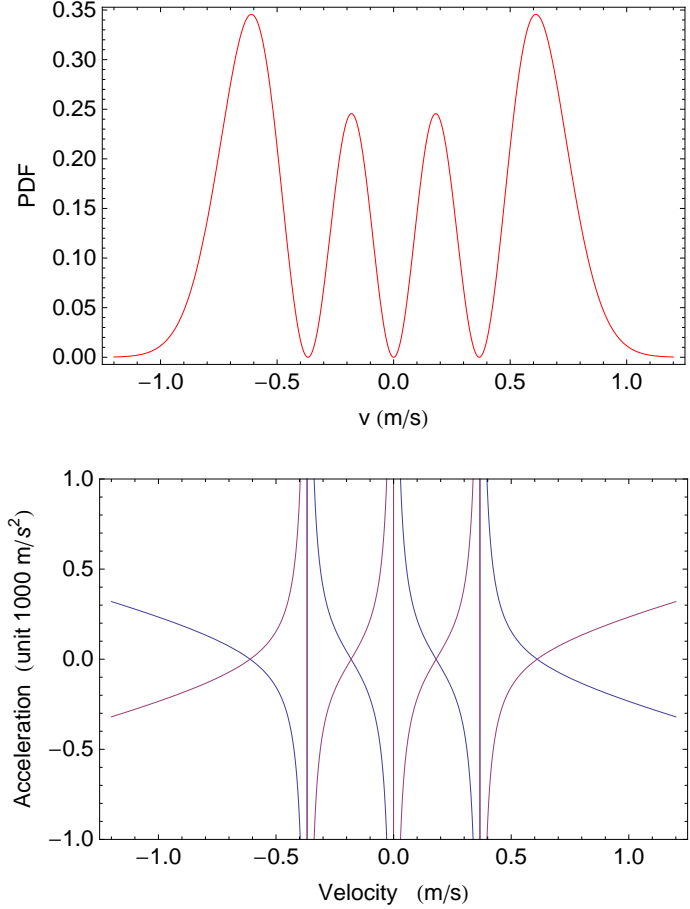} 
\caption{\small{Top figure: PDF of velocity for a macroquantum harmonic oscillator, in the excited state $n=3$, for $v_0=0.3$ m/s. Down figure: accelerations $A_q = \pm D_v \, \d_v \ln P_v$,  expected for the PDF of velocity $P_v(v)$ given in the top figure, for $\Dv=15$. The acceleration becomes divergent on the null minima of the velocity PDF (see top figure).}}
\label{Aq-n3}
\end{center}
\end{figure}

The main implications of this new approach that one can theoretically predict are :

\begin{itemize}

\item The velocity PDF is expected to be locally given by the square of the modulus of a complex `wave function', $P_v=|\psi_v|^2$. 

\item The function $\psi_v$ is a solution of a Schr\"odinger-type equation. It can have any sign and is generally expected to jump from one sign to the other at some values $v_i$.

\item We therefore expect the occurence of velocity values for which $P_v(v_i)=0$. In most cases, one expects $\psi_v\propto \pm(v-v_i)$ around these values, so that $P_v(v) \propto (v-v_i)^2$. A typical example of this behavior is given by harmonic oscillator solutions of the Schr\"odinger equation (see Fig~\ref{Aq-n3}).

\item The new acceleration component $A_q= \pm \Dv\, \d_v \ln P_v$ becomes divergent around these values $v_i$, i.e. $A_q \to ± \infty$ when $v \to v_i$ (Fig~\ref{Aq-n3}). We have suggested \cite{Nottale2014b}  that these divergences are the cause for the large tails of the acceleration PDF in a turbulent fluid. We shall show in the following (Sec.~\ref{PDFofA}) that experimental laboratory data \cite{Voth2002, Mordant2004B, Mordant2001T, Mordant2003} can be precisely fitted by the analytical law that we deduce for the PDF from the expression of $A_q$ (see Fig.~\ref{DoubleCorBoden}).

\item Calm zones. Since the large tails of the acceleration PDF are now accounted for by the new acceleration component $A_q$, the residual stochastic fluctuation is now Gaussian in our model. Therefore, we expect the calm intermittent periods, which correspond to $A_q \approx 0$ and then are given by the mere stochastic background fluctuation, to be Gaussian. This is confirmed by analysis of experimental data (see Fig.~\ref{Gauss}).

\item Numerical simulations. Some analytical results can be obtained from the mere expression of $A_q$, such as the PDF of acceleration (see \cite{Nottale2014b} and the following Sec.~\ref{PDFofA}). One can extend the analysis by performing numerical simulations integrating the system of equations \ref{system} (this is similar to previous simulations performed in $x$-space in the context of standard quantum mechanics \cite{Hermann1997, Nottale2011}). These simulations are used, in particular, to deduce from the Schr\"odinger/$A_q$ process the expected correlation function of the acceleration modulus (see Figs.~\ref{Radetau} and \ref{Rlna}) and the exponents of structure functions (Fig.~\ref{Exponents}), which are in good agreement with the experimentally observed ones.

\end{itemize}

\section{Experimental validations of theoretical predictions}
\label{evote}

Let us review the main experimental validations and proofs of the theoretical predictions made in this new model of turbulence, by giving some typical examples from analysis of experimental data. Detailed explanations of the data analysis leading to these results and more complete works will be presented in forthcoming publications.

\subsection{Empty zones and non-Gaussianity of local velocity PDF}
\begin{figure}[!htp]
\begin{center}
\includegraphics[width=16cm]{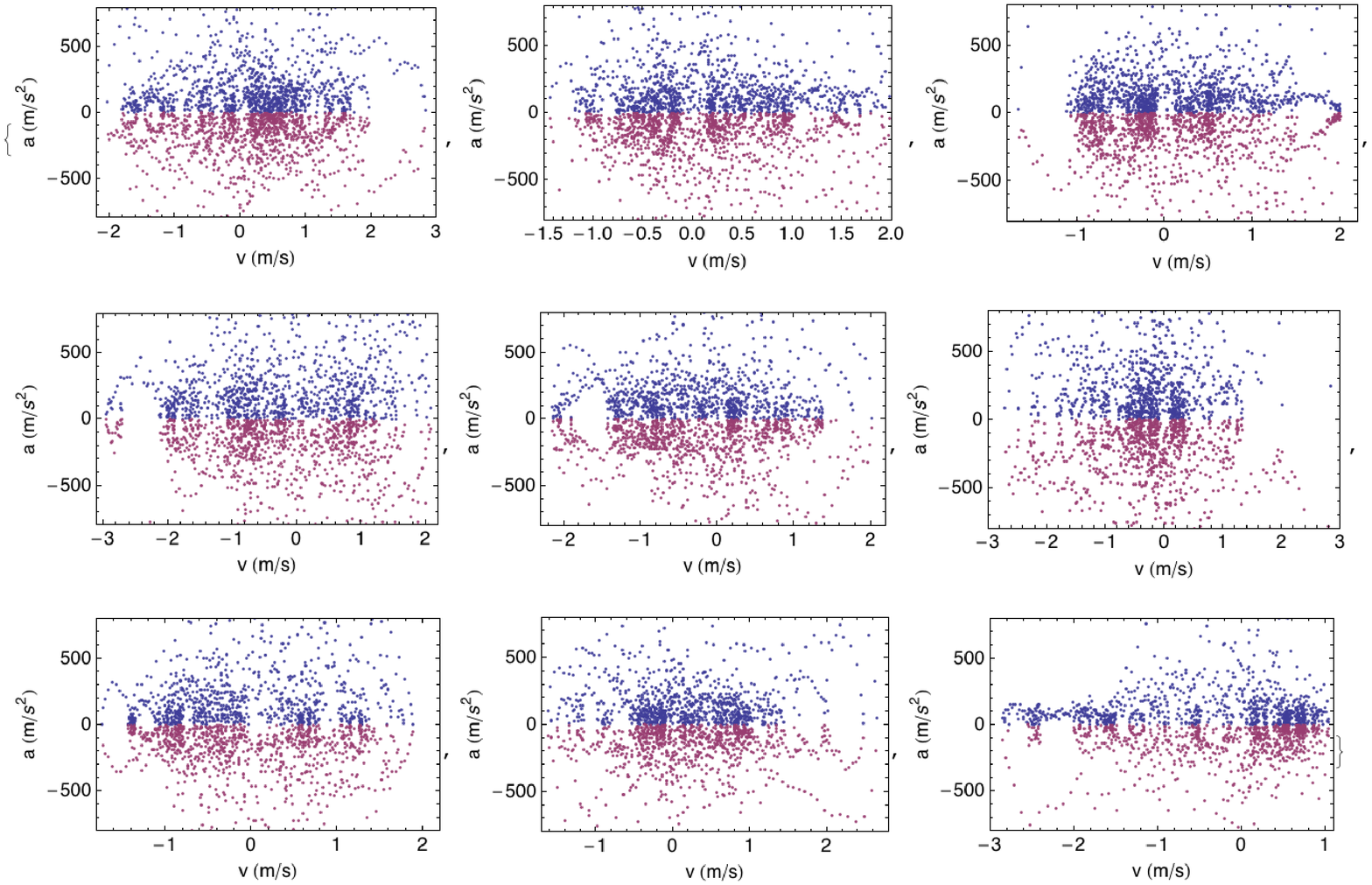} 
\caption{\small{Examples of ``phase" diagram $(v, \; a)$ for nine long segments of Mordant's experiment man290501 (respectively segments 1135, 2578, 3030, 85, 3439, 8292, 9508, 6353, 3500, whose length vary from 2218 to 1729 $t_u$, i.e. $\approx 1700$ to 1300 $\tau_\eta$ and $\approx 15$ to 12 $T_L$). One can easily check that the existence of almost empty minima in the velocity distribution for $|a| <\approx \sigma_a=280$ m/s$^2$ is a systematic property of these Lagrangian segments, although the position of these minima vary, as expected, from one segment to the other. On the contrary, the velocity PDFs for $|a| >\approx \sigma_a$ remains smooth and close to Gaussian (see Fig.~\ref{vat}).}}
\label{neufsegs}
\end{center}
\end{figure}

Some examples of experimental $(v, \; a)$ `phase' diagrams are given in Fig.~\ref{neufsegs}. They very clearly support the systematic existence of empty zones in the ``local" velocity PDF. We call here ``local"  the PDF derived from one unique Lagrangian trajectory or trajectory segment of a fluid particle, instead of the PDF of the full Lagrangian field constructed from adding the data from a large number of different segments. Indeed, this mixing of different trajectories would smooth out the Schr\"odinger-type structures that we expect to be manifested, since the oscillator potential in the Schr\"odinger equation is given by a large scale eddy which is expected to evolve with time and/or to jump to a different one.

Note also that these structures appear only for $a< \approx \sigma_a$, as expected from the existence of a `macroquantum'-classical transition around $a=\sigma_a$ (Sec.\ref{Qcltrans}).  Then the values of the zero minima $v_i$ of $P_v$ change from one segment to the other. Therefore they are not observable through the usual method of analysis, which consists of accumulating a large number of velocity values (e.g., $3 \times 10^6$ in Mordant's experiment man290501 \cite{Mordant2001T, Mordant2004}, more than $10^8$ in some recent experiments \cite{Mordant2004B}). Such a sum will clearly destroy the minima by phase mixing (see Fig.~\ref{neufsegs}) and lead to a final strict Gaussian distribution, as expected from the central limit theorem and as experimentally observed \cite{Mordant2001,Voth2002}.

\subsection{Direct validation of the new acceleration component $A_q$}

\begin{figure}[!ht]
\begin{center}
 \includegraphics[width=10cm]{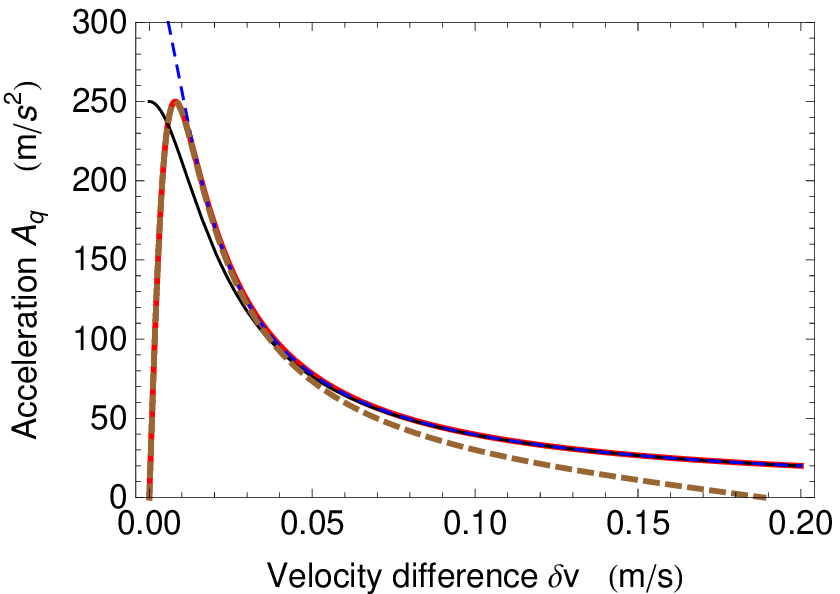}  
\caption{\small{Various models for the magnitude of the new acceleration component, $|A_q|(\delta v)=\Dv \: \d \ln P_v/\d(\delta v)$, in function of the distance $\delta v$ to the nearest zero minimum of $P_v$. The dominant behavior is $|A_q| \propto \delta v^{-1}$, plus asymptotic possible corrections depending on the model (see text).}}
\label{AqPredict}
\end{center}
\end{figure}

\begin{figure}[!ht]
\begin{center}
 \includegraphics[width=10cm]{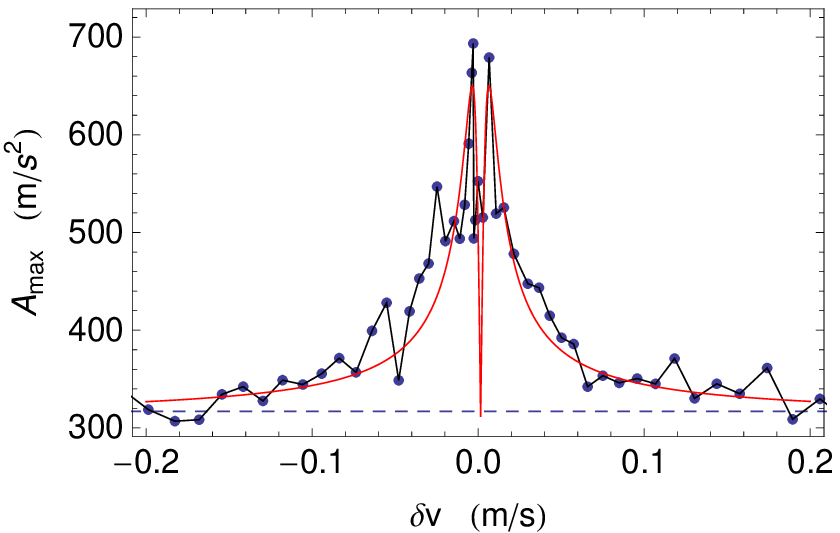} \\  
\caption{\small{Experimental proof of the new acceleration contribution $A_q(\delta v)$. The maximal acceleration in arches is plotted against the distance (in $v$-space) to the closest minimum of $P^q_v(v)$ (points, Mordant's 2D experimental data). It is compared to the theoretically expected relation $A_q(\delta v)$, which is $\sim 1/\delta v$ plus a cut-off toward the origin (Eq.~\ref{Aq2}, red curve, $w=0.0055$ m/s).}}
\label{AqExper1}
\end{center}
\end{figure}

The new acceleration contribution $A_q=\pm \Dv \: \d_v \ln P^q_v(v)$, where $P^q_v(v)$ is the velocity PDF for $|a|< \sigma_a$, can be directly obtained from the local velocity PDF $P^q_v(v)$. This PDF can be established from an histogram of velocities on a given trajectory. The values of $A_q$ depends on the distance of $v$ to the nearest zero minimum $v_i$ of $P^q_v(v)$. Setting $\delta v=v-v_i$, one may write to lowest  order approximation $P^q_v(\delta v) \sim \delta v^2$, so that $A_q=2 \Dv / \delta v$ (blue dashed curve in Fig.~\ref{AqPredict}). The minimum is expected to be not strictly null  in real data, so that one may write $P^q_v(\delta v)=P_0 [1 + (\delta v/w)^2]$ around the minimum. Therefore one obtains an improved formula (red curve in Fig.~\ref{AqPredict})
\beq
A_q(\delta v)=\frac{2 \Dv \delta v}{w^2 + \delta v^2}.
\label{Aq2}
\eeq
The effect of the small threshold $w$ is to cut-off the divergence in $\delta v=0$. In this model, we expect the value of $A_q$ to go down to $0$ in the center of the $P_v$ minimum at $\delta v=0$. A cut-off of the divergence can also be directly inserted in the acceleration by writing $A_q= 2 \Dv/(\sqrt{w^2+\delta v^2}$. In this expression the acceleration no longer vanishes at $\delta v=0$ while it shares the same behavior as the previous ones at large $\delta v$'s (black curve in Fig.~\ref{AqPredict}).

These expressions remains somewhat unrealistic, since they do not take the local maxima of $P_v$ into account. This can be locally modeled by writing 
\beq
P_v = P_0 \l[1 + \l(\frac{v_1}{w}\r)^2 \sin^2 \frac{\delta v}{v_1}\r],
\eeq 
leading to the expression (dashed brown curve in Fig.~\ref{AqPredict}):

\beq
A_q(\delta v)=2 \Dv \frac{ v_1 \cos(\delta v/v_1) \sin(\delta v/v_1)}{w^2 + v_1^2 \sin^2(\delta v/v_1)}.
\eeq
In this case the new acceleration contribution vanishes at a distance $\pi v_1/2$ of the $P_v$ minimum.

The existence of this new acceleration component can therefore be directly checked in Mordant's \cite{Mordant2001T, Mordant2003} experimental data. Actually, $A_q(v)$ is, by construction, defined at the smallest scale of the inertial zone, which is larger than the dissipative scale (due to the transition and, in Mordant's data, to the particle size). Therefore its effect cannot be looked for directly on the individual data points (measured at time-scale $\tau_u=0.7\: \tau_\eta$), but instead on the whole arches themselves. For each segment, we identify the zero minima $v_i$ in the velocity PDF. Then we look for the maximal acceleration of the arches in function of the distance $\delta v$ between the closest $P_v$ minimum and the arch first points (we take the closest of its first three points).  

The result is given in Fig.~\ref{AqExper1}. Its successful comparison with the theoretically expected dependence of $A_q$ in function of $\delta v$ strongly supports the $A_q=\pm \Dv \: \d_v \ln P_v$ process. The final variance balance shows that we have captured most of the acceleration (i.e., of the forces) contributions. For example, in Mordant's 2D data, the standard deviation of the residual accelerations in calm zones is $\sigma_{A\xi} \approx 110-130$ m.s$^{-2}$ while $\sigma_{Aq} \approx 260-290$ m.s$^{-2}$ and $\sigma_a=330$ m.s$^{-2}$. Therefore the basic Gaussian stochastic fluctuations contribute for $10\%-15\%$ of the variance and $A_q$ for $65\%-75\%$. Then there is still probably a small missing component which contributes to $10\%-25\%$ of the variance. This is not unexpected, since we have neglected the  contribution to the acceleration that comes from the phase of the wave function. The approximation of a negligible phase contribution was supported by a description of the largest eddies in the cascade in terms of harmonic oscillator potentials, which leads to real solutions of the Schr\"odinger equation. But it is clear that this cannot be strictly correct, so that another small contribution is naturally expected, which is however difficult to estimate directly.

\subsection{Intermittency of acceleration}

\begin{figure}[!ht]
\begin{center}
\begin{tabular}{cc}
 \includegraphics[width=5.4cm]{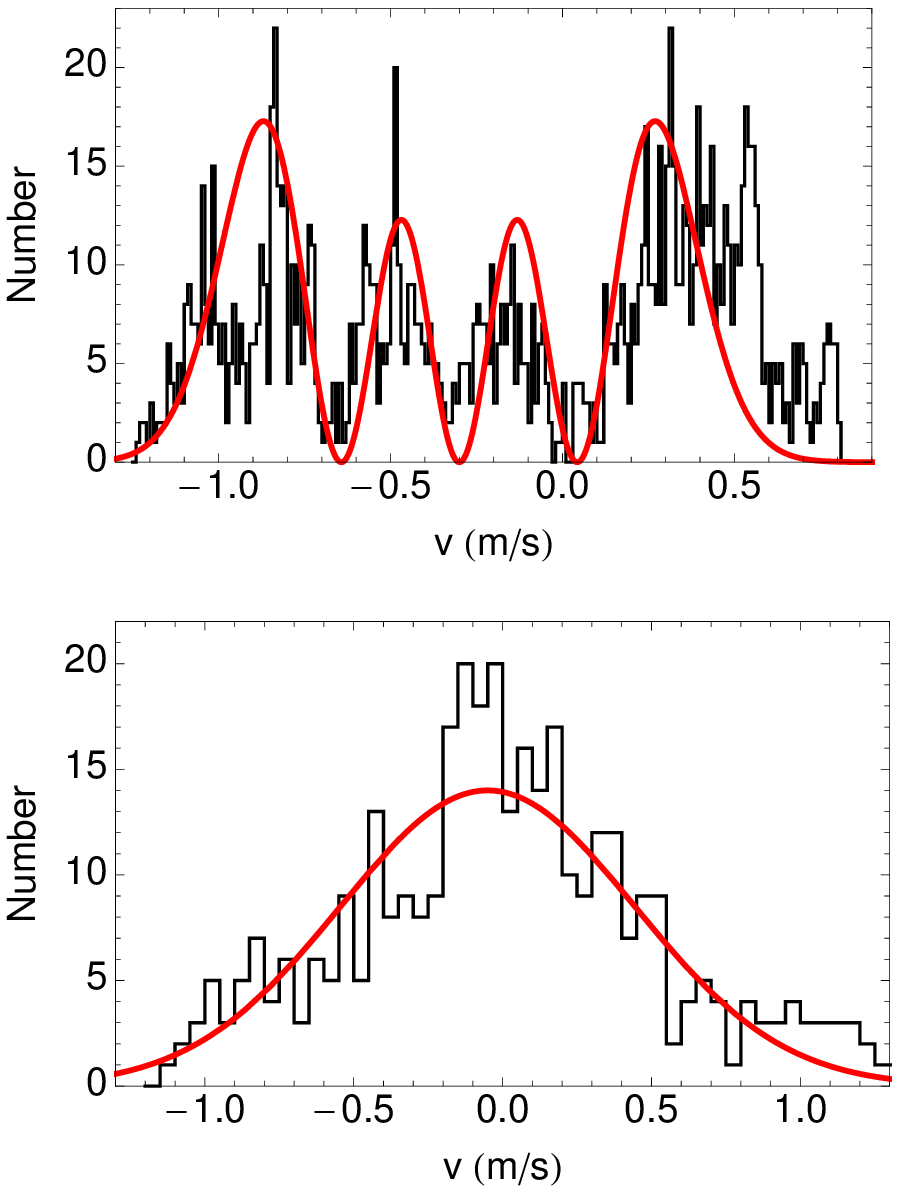} & 
 \includegraphics[width=10cm]{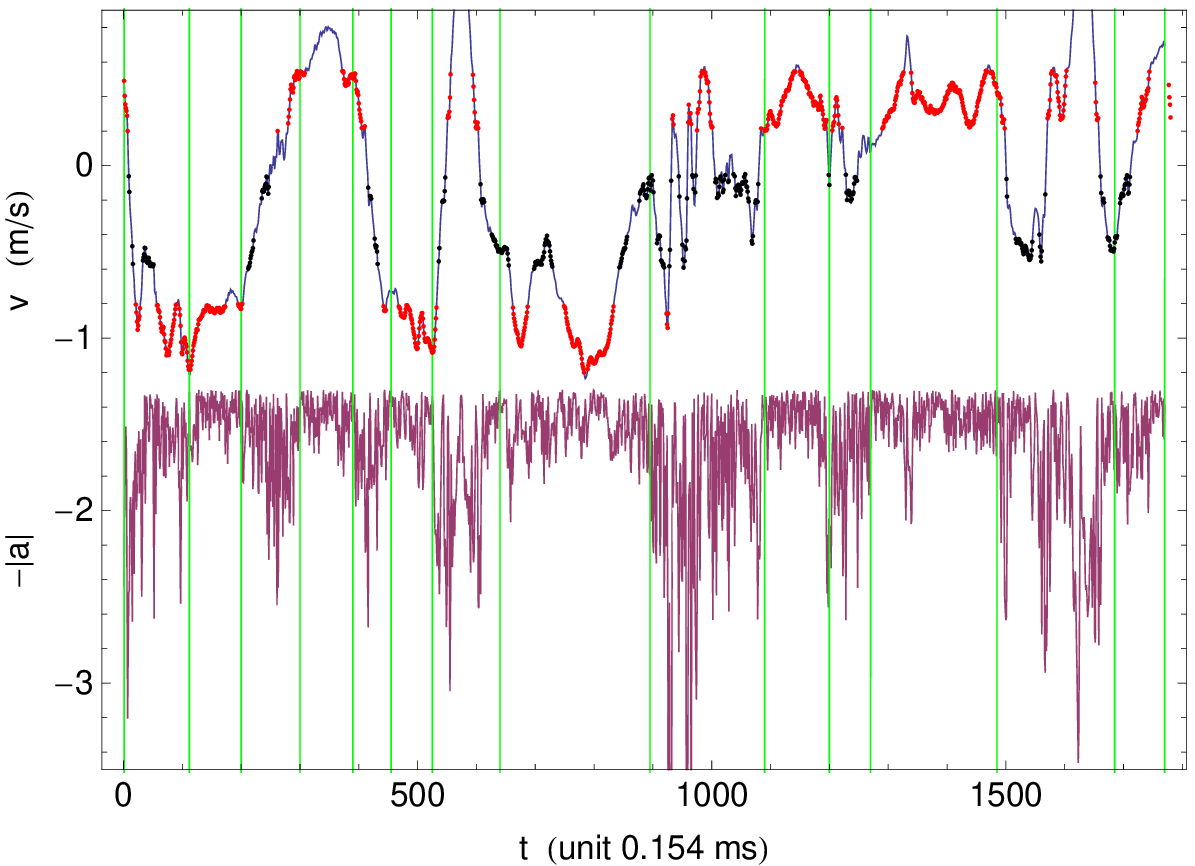} 
\end{tabular}
\caption{\small{Left figure: observed PDF of velocity for the sub-sample [1-1800] of Seg3398 of Mordant's experiment $\#3$ (man290501). Top left figure: observed PDF of velocities for $|a|<\sigma_a$ (black histogram), compared with a solution of the macroscopic Schr\"odinger equation for a quantized harmonic oscillator with $n=3$ (red curve). External peaks not accounted by this theoretical PDF just fit the corresponding classical oscillator. Down left figure: observed PDF of velocities for $|a|>\sigma_a$ (black histogram), compared with a Gaussian of standard deviation $0.7$ m/s (red curve).  Right figure: detailed analysis of the evolution of velocity and acceleration with time for the same sub-segment. The top curve shows the time evolution of velocity $v(t)$. The bottom curve shows the time evolution of the acceleration magnitude ($|a(t)|$ (reversed, varying here from 0 to 1000 m/s$^2$). We have underlined in red the points lying in the main external peaks of the local PDF of velocity and in black those lying in the secondary internal peaks (top left figure). The vertical green lines mark the limit between the calm periods of acceleration ($<\!\sigma_a \!> \approx 90$ m/s$^2$) and the intermittent bursts ($<\!\sigma_a\! > \approx 330$ m/s$^2$).}}
\label{vat}
\end{center}
\end{figure}

In a turbulent fluid, the accelerations show, in function of time, an alternance of calm periods followed by multi-scale bursts of large accelerations (see Fig.~\ref{ada3398}). This behavior is a direct view of the intermittency of the turbulent flow (which is often characterized by non-K41 exponents of structure functions, see below). 

\begin{figure}[!ht]
\begin{center}
\includegraphics[width=12cm]{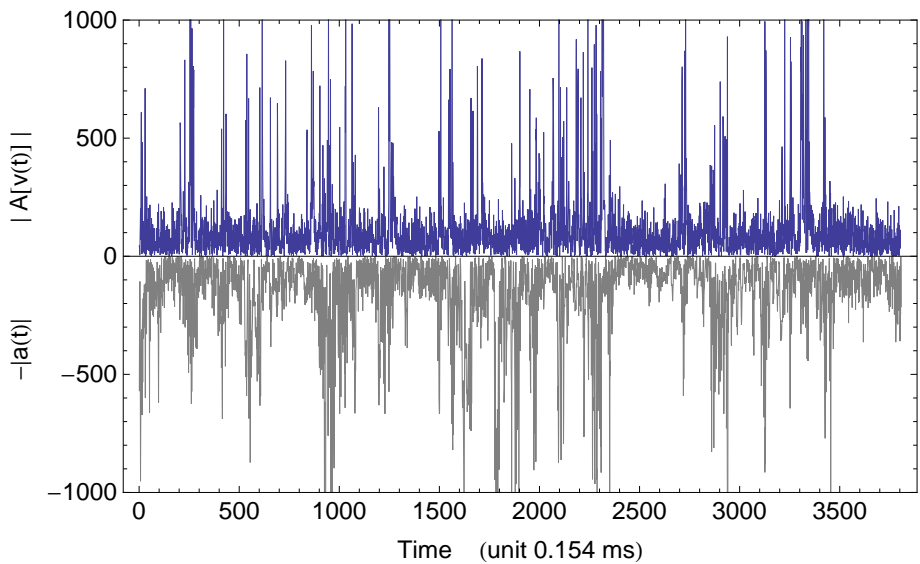} 
\caption{\small{ Comparison between the predicted value of $|A[v(t)]|$ (blue curve, top figure, see text) with the experimentally observed evolution of the acceleration magnitude $|a(t)|$ for the whole segment Seg3398 (gray curve, reversed, down figure). }}
\label{AdeVdeT}
\end{center}
\end{figure}

The new acceleration contribution $A_q[v(t)]=\pm \Dv \, \d_v \ln P_v[v](t)$ yields a detailed explanation and account of this behavior (see Figs.~\ref{vat} and \ref{AdeVdeT}). We give in Fig~\ref{vat} an example of time evolution of velocity and acceleration for a segment of Mordant's data. This segment shows, as predicted, well defined peaks and almost zero minima of the velocity PDF for $a<\sigma_a$, while the PDF is close to Gaussian for $a>\sigma_a$ (left figure). This comes in support of a macroquantum to classical transition around $\sigma_a$.

Moreover, it is clearly seen in this figure that the calm periods correspond to the velocity oscillating into the main probability peaks, while the acceleration bursts result from jumps between these peaks, i.e., from the passage through the minima $P_v(v)\approx 0$ where the new acceleration component $A_q= \Dv \, \d_v \ln P_v$ is expected to diverge.

In addition, we have performed a simulation in which an analytical velocity PDF $P_v(v)$ is given by an harmonic oscillator solution of the Schr\"odinger equation which fits the observed PDF (for $|a|<\sigma_a$) in a long segment of Mordant's experiment (left figure \ref{vat}). Then we compute an analytical expression  $A_q(v)= \Dv \, \d_v \ln P_v$ from this theoretical $P_v(v)$, assuming $\Dv=$cst. Finally we follow the values $v(t)$ experimentally observed for this segment, and for each velocity at each time, we compute $A_q[v(t)]$. According to our model (Eqs.~\ref{system}), we add a random Gaussian fluctuation $A_\xi(t)$ (of standard deviation only one third of that of $A_q$). We have neglected here the Langevin term, which contributes very weakly to the global variance. The result is given in Fig.~\ref{AdeVdeT} and compared to the observed time evolution of acceleration in this segment. The agreement between the two variations is striking (the inter-correlation between the observed and predicted accelerations is at the 10 $\sigma$ level).

\subsection{PDF of acceleration}
\label{PDFofA}
The form of the new contribution $A_q$, which diverges in the minima of $P_v$, yields a natural explanation for the very large values observed for accelerations in turbulent fluids \cite{LaPorta2001}. Moreover, one of the main consequences of the scale relativity model of turbulence is its ability to predict in a detailed way the shape of  the acceleration PDF, including in particular its large tails. 

\begin{figure}[!ht]
\begin{center}
\includegraphics[width=10cm]{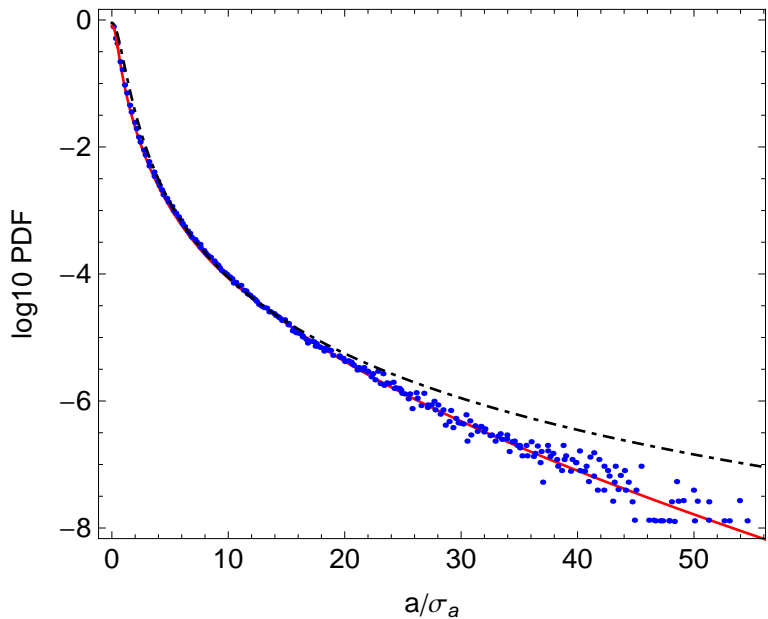} 
\caption{\small{Comparison between the observed acceleration PDF in Bodenschatz et al data \cite{Mordant2004B} ($R_\lambda=690$, $10^8$ values up to $\sim55$ sigmas, blue points), the $(1+(a/\sigma_a)^2)^{-2}$ law expected as a first approximation (dot-dashed black curve) and the PDF corrected for both small and large accelerations  (red continuous curve). The corrected curve perfectly fits the data within experimental uncertainties.}}
\label{DoubleCorBoden}
\end{center}
\end{figure}

The first  step consists of deriving the PDF expected from the new acceleration  $A_q=\pm \Dv\, \d_v \ln P_v$, which is now the dominant contribution. The large tails of the PDF are simply explained by the fact that $P_v= |\psi_v|^2$ and that the modulus of the wave function $\psi_v$ is expected to oscillate between positive and negative values. When it crosses $\psi_v=0$ at some velocity $v_i$, one expects in most cases $\psi_v  \propto \pm (v-v_i)$, and therefore $P_v  \propto  (v-v_i)^2$. Around these zeros, one gets  $A_q=\pm2 \Dv/(v-v_i)$, which is divergent around $v=v_i$. This provides a simple explanation for the experimentally observed very large tails of the acceleration PDF. From this expression of $A_q$, it is easy to deduce the corresponding acceleration PDF in the tails  \cite{Nottale2014b}:
\beq
P_a(a) \propto \frac{1}{a^4}.
\eeq
One can easily improve this model by now describing both a local minimum in $\sim v^2$ and a nearby maximum of $P_v$ in terms of a locally sinus function,  $P_v(v) \propto \cos^2(v/v_0)$.  Setting $\sigma_a=2 \Dv/v_0$, one obtains the acceleration PDF:
\beq
P_a(a)= \frac{2}{\pi {\sigma_a}}\; \frac{1}{[1+(a/{\sigma_a})^2]^2}.
\label{PDvFbox}
\eeq
Note that such a law has already been proposed by Beck \cite{Beck2001,Beck2001a,Beck2005} under the assumption of Tsallis statistics.  In our approach it is predicted from the dynamics, but this is only a lowest order approximation, since this result is obtained from the mere behavior of the inertial range ($\delta v^2 \sim \delta t$). Despite this, it fits the experimental PDF very nicely up to about $15$ sigmas (see Fig.~\ref{DoubleCorBoden}).

\begin{figure}[!ht]
\begin{center}
\includegraphics[width=14cm]{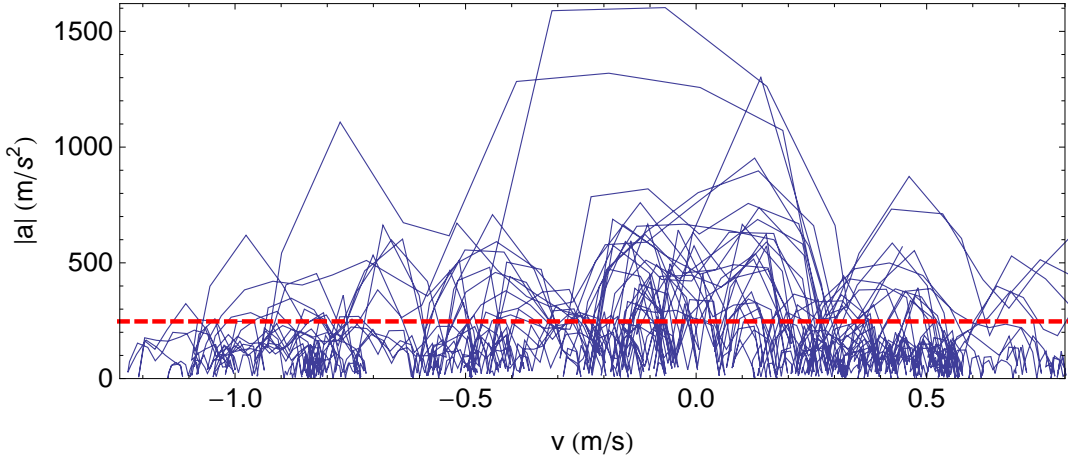} 
\caption{\small{The ``phase" diagram $(v,\, |a|)$ in velocity space for the first part of seg3398 of Mordant experiment man290501 ($t= \tau_u$ to $1770 \; \tau_u$). The straight lines connect the measurement points which are separated by $\tau_u=1/6500$ s. The dashed line gives the local  standard deviation of acceleration $\sigma_a=245$ m/s$^2$ for this segment. The holes in the velocity PDF are clearly apparent for $a< \sigma_a$, while the large values of the acceleration ($a \gg \sigma_a$) evolve along ``archs" which amplify the effect of the new acceleration component $A_q$.}}
\label{Arch1}
\end{center}
\end{figure}

Therefore, one should also account for the dissipative scales, which  cut-off the acceleration divergence toward small time-scales. For scales smaller than a few $ \tau_\eta$'s, the scaling law $\delta v \sim \delta t^{1/2}$ is no longer valid and one comes back to a standard differential $\delta v \sim \delta t$. 
Therefore, toward small time-scales ($\tau< \approx \tau_\eta$), the new acceleration contribution is no longer given by  $A_q=\pm \Dv \: \d_v \ln P_v$ with $\Dv$ constant. It tends to 0 when $\tau \to 0$ and becomes smaller than the stochastic fluctuation which is itself no longer scale-dependent. Therefore, at the limit $\tau \to 0$, one may write $\sigma_a^2=2 \Dv/ \tau$, where $\sigma_a^2$ is the asymptotic value of the acceleration variance, so that one obtains: 
\beq
A_q(\tau,v)=\frac{1}{2} \sigma_a^2 \: \tau \; \d_v \ln P_v(v).
\label{smallscales}
\eeq
One therefore expects the real acceleration PDF to be lower than the purely inertial $a^{-4}$ law.

There are two consequences of this process. One direct consequence concerns the central part of the PDF (``small" accelerations, $|a|<10 \, \sigma_a$ in Bodenschatz et al. data \cite{Mordant2004B}, $|a|<4.5 \, \sigma_a$ in Mordant's data due to the large particle size). By using the small scale expression of $A_q$, Eq.~\ref{smallscales}, one finds a corrected law,
\beq
P_a(A)=\int_0^\infty  \exp \l(-2 \, \frac{A}{t}  \r) \, \frac{(1+t^2)^{-2}}{t} \, dt,
\eeq
which can be integrated in terms of special functions \cite{NottaleLehner} and which perfectly fits the experimental data in the central part of the PDF (see Fig.~\ref{DoubleCorBoden}).  Beyond a few $\sigma_a$'s, it becomes identical to the $a^{-4}$ law. Therefore this effect does not impact the very large tails of the PDF.

\begin{figure}[!ht]
\begin{center}
\includegraphics[width=8cm]{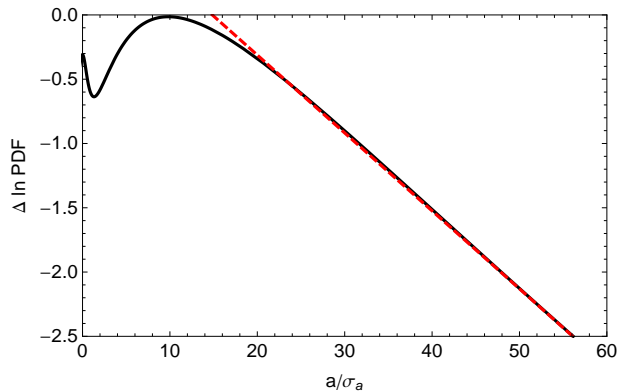} 
\caption{\small{Ratio between the Voth et al \cite{Voth2002} phenomenological formula (that fits the experimental acceleration PDF) and the $a^{-4}$ law theoretically expected as a first order approximation (see text). The black curve is the difference betwen the logarithms of the two PDFs. It is compared to an exponential cut-off (red dashed line).}}
\label{Pa-difference}
\end{center}
\end{figure}

\begin{figure}[!ht]
\begin{center}
\begin{tabular}{cc}
 \includegraphics[width=8.3cm]{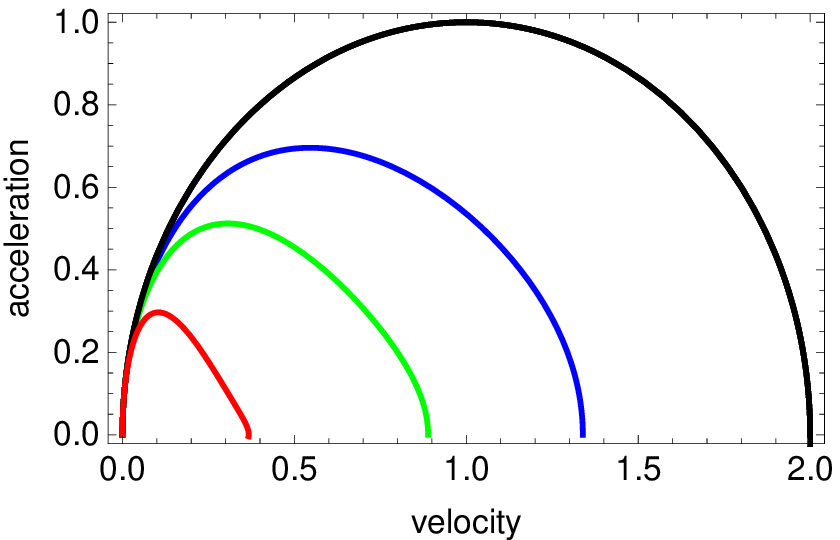} &
 \includegraphics[width=6.7cm]{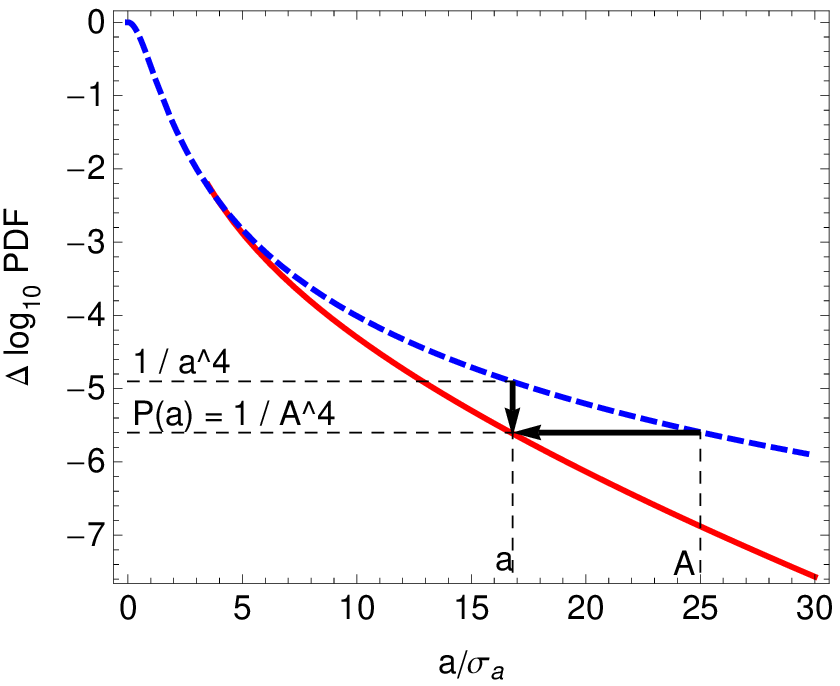} 
\end{tabular}
\caption{\small{Illustration of the mechanism of exponential cut-off to the $a^{-4}$ law (first order approximation of the acceleration PDF). Left  figure: decrease of the maximal acceleration reached for similar initial conditions (the acceleration scale is here arbitrary), in function of the damping parameter $\chi$: from top to down, $\chi=0$ (harmonic oscillator, highest black curve), $\chi=0.2$ (blue curve), $\chi=0.4$ (green curve) and $\chi=0.8$ (lowest red curve). Right figure: an acceleration which should have been $A$ with an $A^{-4}$ PDF according to the inertial regime is decreased to $a= \rho(A) \, A$ due to the damping mechanism of the left figure. Since $\chi$ is correlated with $A$ (see text), the depletion is larger for higher accelerations.}}
\label{damping}
\end{center}
\end{figure}

But there is a second indirect consequence of the existence of the dissipative transition which now applies to the large tails.  The largest values of the acceleration are obtained at the smallest time-scales, i.e. in the dissipative eddies at sub-Kolmogorov scales. These eddies can be described in terms of damped harmonic oscillators (in $v$-space), which are solutions of the equation:
\beq
\frac{da}{dt}= -\frac{a}{T_a}- \frac{(v-v_0)}{\tau_\omega^2}.
\eeq
One can show \cite{NottaleLehner} that this simple model yields solutions that are very close to the anharmonic oscillators directly obtained from the Navier-Stokes equations at dissipative scales, $dv/dt=\nu \Delta v$. These damped solutions achieve ``arches" (see Figs.~\ref{Arch1} and \ref{damping}) whose shapes are characterized by the parameter 
\beq
\chi=\frac{\tau_a}{4 T_a}=\frac{\tau_\omega}{2\sqrt{4 T_a^2- \tau_\omega^2}}.
\eeq
One finds for the arches the general relation:
\beq
\frac{A_{\rm max} T}{ \pi \Delta V}=K(\chi)=1 + \chi^2 +...,
\eeq
where $A_{\rm max}$ is the maximal acceleration reached on half a period, $T$ the half-period duration and $\Delta V$ the half velocity extension of the arch. Then, at small time-scales close to the dissipative scales, the maximal acceleration $A_{\rm max} \propto A_q \propto \tau$. One identifies the scale variable $\tau$ to $\tau_a$ and  since the damping time $T_a$ does not depend on scale, one finds that  $A_{\rm max} \propto  \tau_a \propto \chi$. This means that the probability distribution of the arch shapes is not constant in function of their amplitude, but instead that their maximal acceleration is correlated with their shape. Larger is the maximal acceleration, larger is $\chi$, i.e., the largest accelerations are obtained in arches (eddies) which are more damped. This expectation has been checked in Mordant's data \cite{Mordant2001T}: one can show that there is a quadratic correlation between the parameter $K \approx 1 +\chi^2$ and the maximal acceleration of an arch $A_{\rm max}$ at the $17\sigma$ level of significance, which supports the theoretical expectation  $A_{\rm max} \propto   \chi$ \cite{NottaleLehner}.

One naturally expects an exponential cut-off  to the $a^{-4}$ purely inertial law from such a relative damping mechanism. This is supported by the experimentally observed PDF (see Fig.~\ref{Pa-difference}).
 
Therefore, an acceleration which should have been $A$  with an $A^{-4}$ PDF without this effect is reduced to a smaller acceleration $a$ by a damping factor $\rho(a)=a/A(a)$  (see Fig.~\ref{damping}). This damping factor can be established from the analytical expressions of the arches obtained from solving the viscous Navier-Stokes equation at dissipative scales, $dv/dt=\nu \Delta v$ \cite{NottaleLehner}.

As expected, this process results in an exponential cut-off to the PDF tails, such that the corrected PDF is found to be given, for Bodenschatz et al data, by
\beq
P_a(a) \propto \frac{e^{-2 \sqrt{1+ ( {a}/{26.7 \sigma_a})^2}}}{[1+(a/\sigma_a)^2]^2},
\eeq
where the numerical coefficient results from a fit of the experimental data \cite{NottaleLehner}. 

The corresponding PDF for Mordant data \cite{Mordant2001T} is obtained from a simple dilation of the acceleration values, $a_B \approx 2.3 \: a_M$ which accounts for the particle size effect (amounting to an effective Kolmogorov scale $\tau_{\eta {\rm eff}} \approx 5 \, \tau_\eta \approx 2.3^2 \tau_\eta$). Note that, in the observed range ($|a|>\approx 15 \, \sigma_a$ in Bodenschatz et al data), this exponential cut-off is indistinguishable from the simple following law, depending on only one free parameter $a_0$:
\beq
P_a(a) \propto \l( \frac{1}{a}-\frac{1}{a_0} \r)^4,
\eeq
where $a$ is normalized to $\sigma_a$. The free parameter is fitted to $a_0=44$ for Mordant's data and $a_0=100$ for Bodenschatz et al data (in agreement with the ratio $\approx 2.3$ between the two sets of data).

It has been shown \cite{Voth2002,Mordant2004B} that the experimental acceleration PDFs, once normalized, exhibit universal behavior for high Reynolds numbers ($R_\lambda> \approx 500$) and can be well fitted by a phenomenological stretch exponential formula. Our resulting predicted PDF shows an excellent agreement , with only one fitted parameter, with both this phenomenological fit and the experimental data (see Fig.~\ref{DoubleCorBoden}).

\subsection{Gaussianity of residual fluctuations in calm zones}
\label{gaussianity}

\begin{figure}[!ht]
\begin{center}
\begin{tabular}{cc}
 \includegraphics[height=4.9cm]{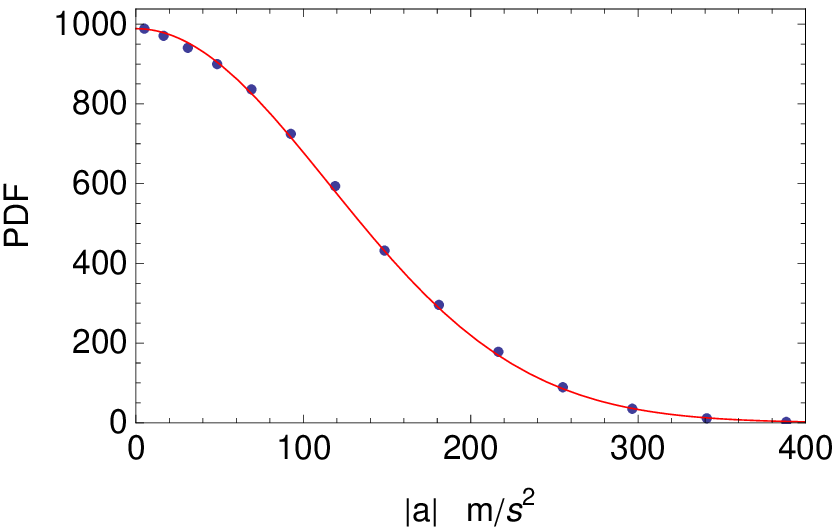} & 
 \includegraphics[height=4.9cm]{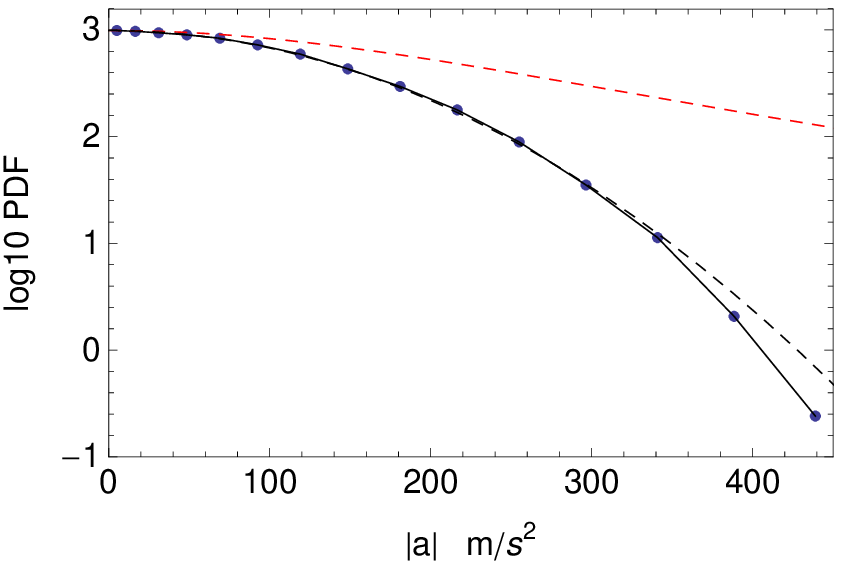} \\  
\end{tabular}
\caption{\small{Left figure: observed PDF of accelerations in the calm zones (points, from Mordant 2D experiment \cite{Mordant2001T}), compared with a Gaussian (red curve). Right figure: logarithm of this PDF, confirming its Gaussian nature (black dashed curve) and the absence of large tails, compared to the $a^{-4}$-like global PDF (red dashed curve).}}
\label{Gauss}
\end{center}
\end{figure}

One of the most radical new predictions of the scale relativity model of turbulence is that the residual fluctuations, beyond the effect of $A_q$, remains Gaussian. This is opposite to the standard description in which, except for the small Langevin term, the whole acceleration is stochastic and should therefore be characterized by the highly non-Gaussian PDF with large tails. 

In our model, the large tails are generated by $A_q$ while the stochastic residual fluctuation $A_\xi$  is expected to be a standard Gaussian Brownian motion with variance $\sigma_{A\xi}^2  \ll \sigma_{a}^2$.

It is easy to check for this property in experimental data, since we expect the acceleration to be reduced to the fluctuation $A_\xi$ when $A_q$ vanishes, and we kwow precisely when this occurs: just when the velocity $v$ is far from the zero minima of $P^q_v$, namely, when the particle trajectory oscillates inside the probability peaks of the velocity PDF, without crossing the null minima (see Fig.~\ref{vat}). This corresponds to the ``calm" zones of the intermittent acceleration (see e.g. Fig.~\ref{ada}). 

This expected correlation is supported by a statistical analysis of Mordant's experimental data : in the 2D data (which allows to correct for the $v=0$ bias present in the 1D data), we find that $85 \%$ of the calm zones are inside the probability peaks of $P^q_{v}(v)$ ($|a|< \sigma_a$).

We give in Fig.~\ref{Gauss} the PDF of acceleration in these calm zones in Mordant's experimental data. It is fully Gaussian with a high level of precision, showing absolutely no non-Gaussian large tail. Its standard deviation is $\sigma_{A\xi}=115$ m.s$^{-2}$ which is 40$\%$ of the full $\sigma_{a}=280$ m.s$^{-2}$. In other words, the residual Gaussian fluctuation contributes to now only $\approx 16 \%$ of the variance and therefore $A_q$ to $\approx 84\%$ or less (since a small missing phase term may also contribute, see Eqs.~\ref{AA} and what follows).

\subsection{Autocorrelation functions of acceleration modulus}

\begin{figure}[!ht]
\begin{center}
\begin{tabular}{cc}
 \includegraphics[width=7.5cm]{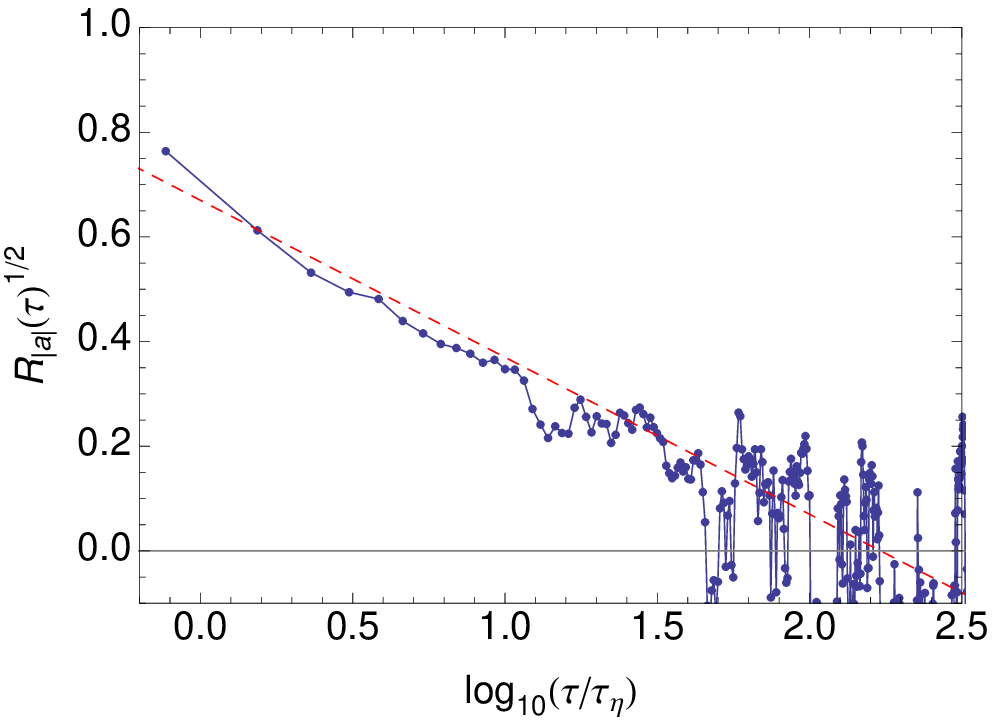} & 
 \includegraphics[width=7.5cm]{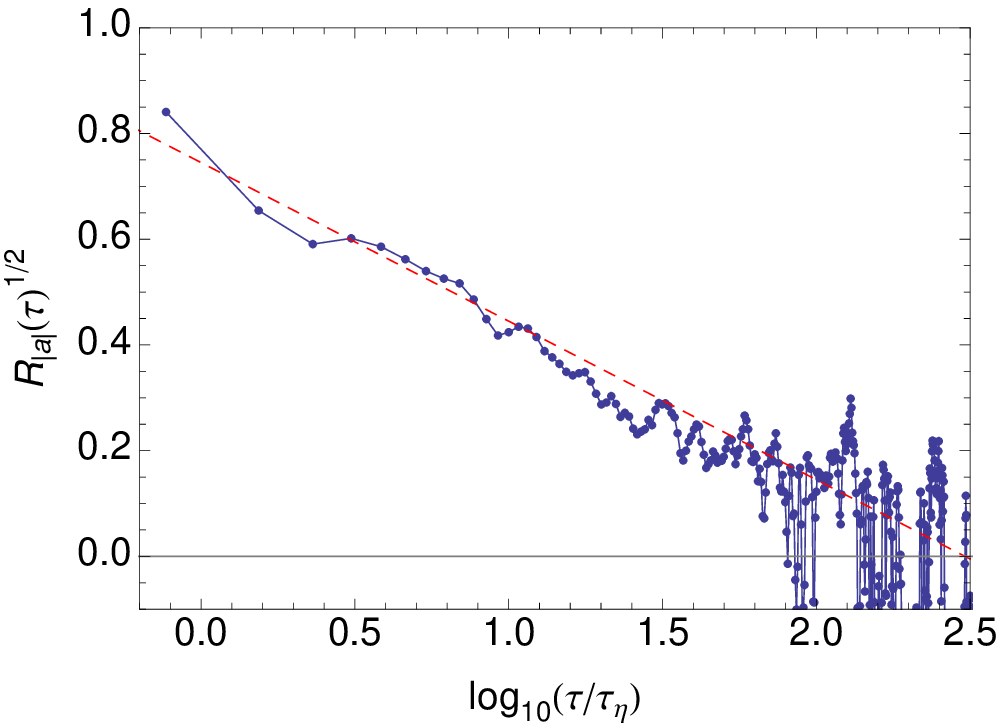} \\
\end{tabular}
\caption{\small{Square-root of the autocorrelation function $R_{|a|}^{1/2}$ of the acceleration magnitude. The left figure is  obtained from a numerical simulation of the acceleration values $A_q[v(t)]$ derived from a velocity PDF given by a $n=3$ quantized harmonic oscillator. It compares very well with the right figure, which is obtained from the experimental values of acceleration for Seg3398. The dashed red lines are linear fits with yield similar slopes in the simulation and in the experimental data.}}
\label{Radetau}
\end{center}
\end{figure}

\begin{figure}[!ht]
\begin{center}
\begin{tabular}{cc}
 \includegraphics[width=7.5cm]{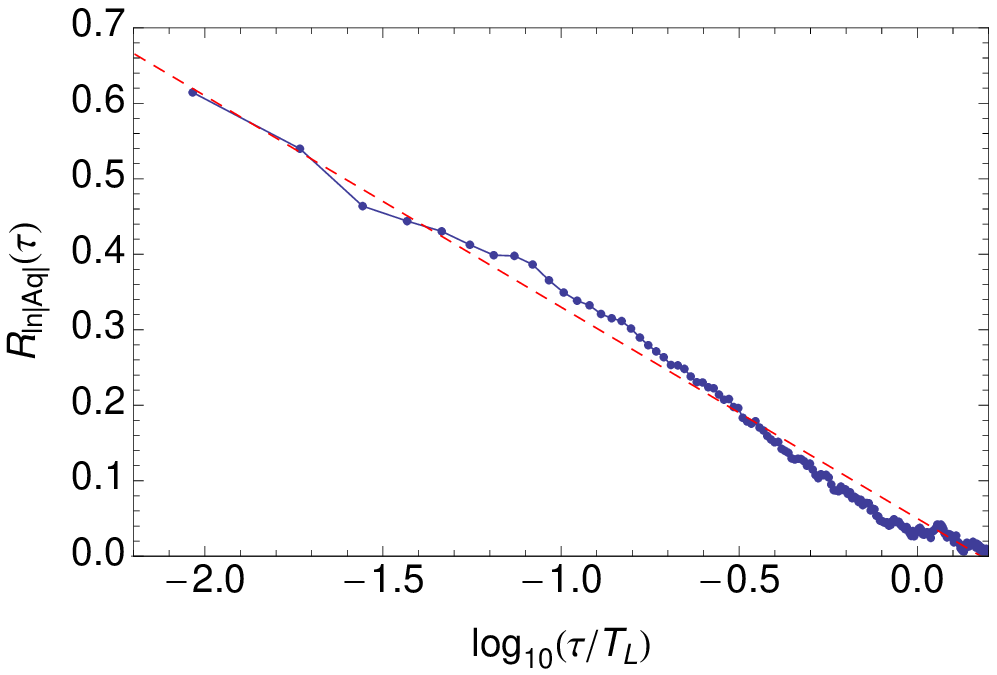} & 
 \includegraphics[width=7.5cm]{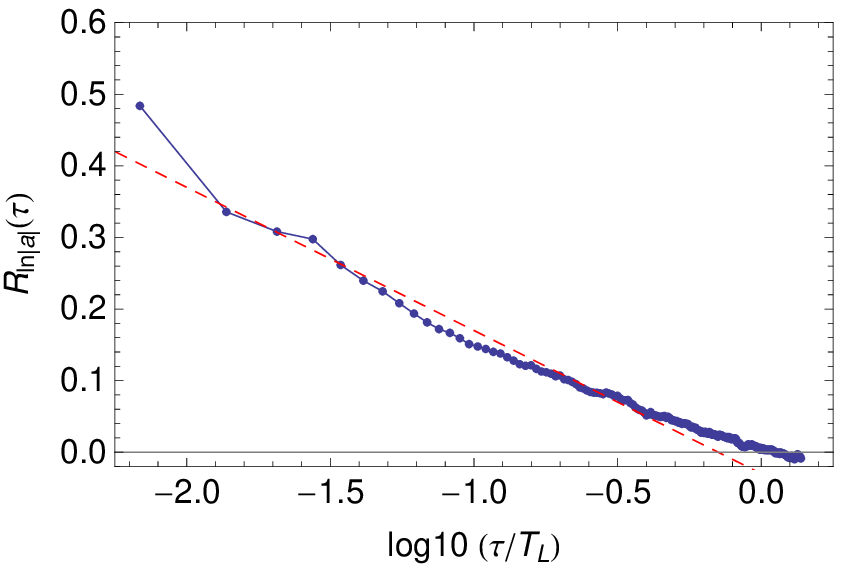} \\
\end{tabular}
\caption{\small{Autocorrelation function $R_{\ln |a|}(\tau)$ of the logarithm of the acceleration magnitude. The left figure is  obtained from a numerical simulation of the acceleration values $A_q[v(t)]$ derived from a velocity PDF given by a $n=1$ quantized harmonic oscillator. It compares well with the right figure, which is obtained from the experimental values of acceleration for Mordant's experiment man290501. The red dashed lines are linear fits which yields similar slopes in the simulation and in experimental data.}}
\label{Rlna}
\end{center}
\end{figure}

One of the most interesting results of recent experimental turbulence studies is the discovery of a linear long-range behavior for the square root of the autocorrelation function $R_{|a|}^{1/2}$ of the acceleration modulus (see right figure~\ref{Radetau}). A similar result holds for the correlation function of its logarithm, $R_{\ln|a|}$ (right figure~\ref{Rlna}) \cite{Mordant2001T, Mordant2003}. This result is used as empirical basis for multifractal random walk models \cite{Arneodo1997, Bacry2001, Castaing1990, Frisch1995} which have allowed to recover, e.g., the values of some of the exponents of structure functions.

These models are just built from the assumption of an autocorrelation of the acceleration magnitude given by \cite{Mordant2003}:
\beq
<\ln |A(t)|\; \ln |A(t+ \tau)|>= - \lambda^2 \,  \ln \frac{\tau}{T_L},
\eeq
where $ \tau= \delta t < T_L$ and $\lambda^2$ is an adjustable parameter.

We have performed numerical simulations of the Schr\"odinger/$A_q$ process (Eqs.~\ref{system}) and then computed $R_{|a|}^{1/2}$ and  $R_{\ln|a|}$ for the accelerations obtained in these simulations. Some typical examples of the results are given in the left parts of  Fig.~\ref{Radetau} and Fig.~\ref{Rlna}. We obtain the same results independently of the chosen value for the quantum number $n$ characterizing the oscillator in the Schr\"odinger equation. This independance is expected since the dominant effect comes from the minima of $P^q_v(v)$ which have a universal parabolic shape.
The simulations look very much like the experimentally observed autocorrelation functions (right figures), with values of the slope close to the observed one $-\lambda^2 \approx -0.12$. This suggests that a universal process may be at play and that it should be possible to get an analytical expression of these autocorrelation functions. 

\begin{figure}[!ht]
\begin{center}
 \includegraphics[width=8cm]{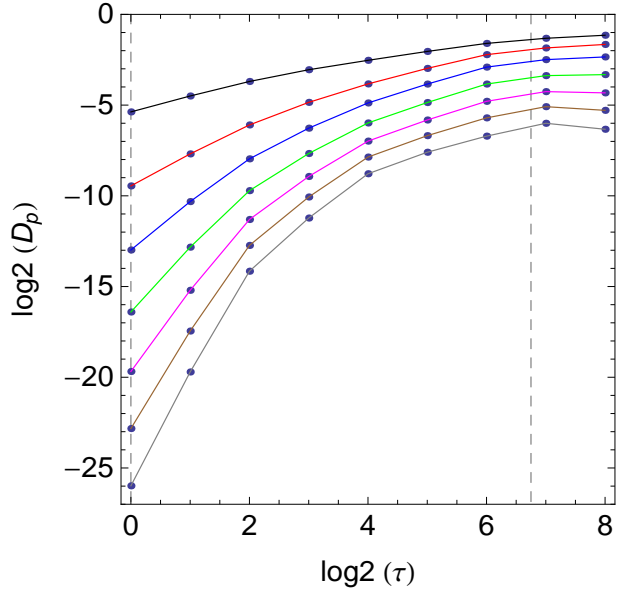} \\  
\caption{\small{Structure functions obtained in a numerical simulation of the $A_q$-Schr\"odinger process, $D_1$ to $D_7$, from top to down. The velocity PDF is solution of a Schr\"odinger equation for an harmonic oscillator potential with $n=3$. The trajectory in the simulation contains $N=80000$ points sampled with time intervals $\tau_\eta$, then smoothed with a filter of width $5 \tau_\eta$ to account for a particle size $250\; \mu$m, as in Mordant's experiment. The simulated structure functions  are in close agreement with those obtained from the experimental data in this experiment \cite{Mordant2001T,Mordant2003}.}}
\label{StructFunct}
\end{center}
\end{figure}

\subsection{Exponents of structure functions}

\begin{figure}[!ht]
\begin{center}
 \includegraphics[width=8cm]{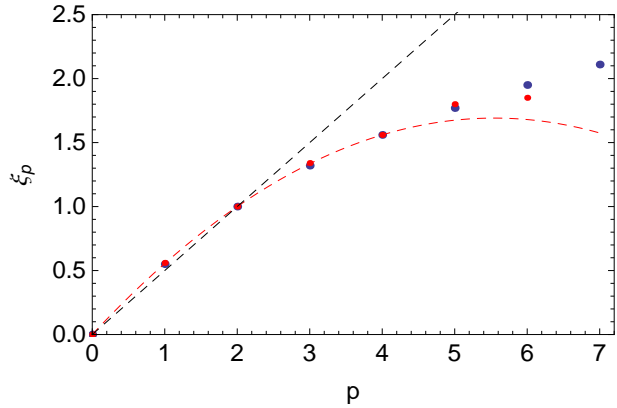} \\  
\caption{\small{The exponents of the structure functions measured in a numerical simulation of the Schr\"odinger-$A_q$ process for a harmonic oscillator potential with $n=3$ (blue points) are compared to those measured in experimental data (\cite{Mordant2001T} red points), to the K41 expectation (black dashed line) and to the exponents derived from a multifractal random walk model (\cite{Bacry2001}, red dashed curve).}}
\label{Exponents}
\end{center}
\end{figure}

One of traditional way of characterizing intermittency is by the difference of observed exponents of the structure functions with respect to their expected K41 values.

We have achieved numerical simulations of the $A_q$-Schr\"odinger process and computed the structure functions for this simulations,
\beq
D_p (\tau)= < |\Delta_\tau v|^p>.
\eeq 

Then we have established their scale dependance (see Fig.~\ref{StructFunct}) and measured the exponents $p$ through Extended Scale Similarity \cite{Benzi1993}, in which one looks for the scale dependence of the structure functions relatively to the second order moment $D_2$, i.e., $D_p(\tau)=D_2(\tau)^{\xi_p}$.

The result is compared in Fig.~\ref{Exponents} to the exponents measured by Mordant in his experimental data \cite{Mordant2001T, Mordant2003}. The experimentally measured exponents are closely reproduced (within uncertainties) by the simulation, even for large values of $p > 4$ for which the various existing models no longer account properly for them \cite{Schmitt2005}.

\section{Discussion and conclusion}
\label{conclusion}

Let us explain in more detail the mechanism underlying the new process. The inertial range and its K41 scaling law $\delta v^2\sim \delta t$ serves as microscopic structure and theory for the establishment of the Schr\"odinger equation. The oscillator potential in this equation therefore corresponds to the largest eddies of the cascade, at the large scale end of the inertial domain. These large eddies remain established during times that may be far larger than the integral time $T_L$ (a full turn of the particle in the eddy lasts $\approx \pi t_L= \pi L/\sigma_v\approx 6-9 \;T_L$ and several turns are possible). This explains why a macroquantum structure may exist on times $\gg T_L$. 

Then the new acceleration component $A_q$ is generated from the solution $P^q_{v}= |\psi|^2$ of this Schr\"odinger equation. It is expected to describe the observed PDF only for $|a| <\approx \sigma_a$, since there is a macroquantum to classical transition around $\sigma_a$.  The experimentally observed PDF is a mixing of macroquantum and classical contributions. We expect these solutions of a Schr\"odinger equation to be manifest only in Lagrangian data, and only  ``locally" (i.e., on a given unique trajectory or segment of trajectory of a fluid element). Indeed, the combination of different Lagrangian trajectories smoothes out the Schr\"odinger structures due to shift of the minima and maxima of $P_v$ for different oscillator potentials describing different eddies. In the same way, the Eulerian data correspond to different trajectories for each measurement, involved in different eddies, thus preventing from manifesting the expected  structures. 

 The scale where $A_q$ is established is the lowest end of the inertial range, just larger than the viscous Kolmogorov scale $\tau_\eta$. At dissipative scales, below $\tau_\eta$, one recovers a standard differential regime $\delta v \sim \delta t$, and the particle trajectories become again deterministic. They are involved in dissipative eddies well described by classical damped or anharmonic oscillators. Although these eddies have the smallest radii and velocities, their accelerations are the largest of the cascade. These ``arches" achieve a kind of amplification to very large accelerations $a \gg \sigma_a$ of the $A_q$ component generated in the macroquantum domain.
 
Let us finally recall the key-points of the new scale relativity approach to turbulence:

(1) Under fractal and non-differentiable conditions, the derivative of the Navier-Stokes equations can be transformed, in the inertial domain, into a Schr\"odinger-type equation;

(2) The velocity PDF, $P_v=|\psi_v|^2$ is obtained from a ``local" solution $\psi_v$ (i.e., concerning one segment of trajectory of a fluid particle) of this macroscopic Schr\"odinger equation (in the sense that the constant which replaces $\hbar$ is a macroscopic diffusion coefficient). This Schr\"odinger regime is valid for $|a|< \approx \sigma_a$, beyond which there is a transition to classical behavior (which a manifestation of the de Broglie transition in $v$-space).

(4) The fact that $P_v$ is the square of the modulus of the ``wave function" $\psi_v$ implies that null minima $P(v_i)=0$ must exist for some values of the velocity. An analysis of Mordant's experimental data has allowed us to show that the velocity PDFs of individual Lagrangian segments are indeed highly non-Gaussian and present in a systematic way empty zones in the ``macroquantum" domain $|a|< \approx \sigma_a$. This quasi-total exclusion of some particular values of the velocity appears as an highly non-classical feature. This non standard behavior is re-inforced by the fact that the velocity PDF remains near to Gaussian when $|a|> \approx \sigma_a$, i.e. beyond the  expected macroquantum to classical transition.

(5) The existence of a new acceleration component $A_q= \pm \Dv \, (\d_v P_v)/P_v$ is deduced from this model. Due to its $1/P_v$ dependence, this new component is divergent at the quasi-null minima $P_v(v_i) \approx 0$. This explains the very large values observed for accelerations in turbulent fluids. This new component has been directly seen and validated in the experimental data.

(6) Moreover, the general behavior of the velocity PDF around its minima, $P_v \sim \delta v^2$, allows one to derive a universal shape for the acceleration PDF under a pure inertial regime, $P_a(a) \sim 1/a^4$. The account for the Kolmogorov dissipative range leads to correct this law in terms of an exponential cut-off at large accelerations. The finally obtained PDF perfectly fits the experimentally observed one with only one free parameter, including its very large tails (which have been measured up to $\approx 55 \, \sigma_a$).

(7) The predicted existence of an alternance of probability peaks and of almost empty minima in the velocity PDF of individual segments also accounts for the intermittency of acceleration in great detail. When the particle velocity oscillates inside one of the probability peaks without crossing a minimum, the $A_q$ contribution remains small and only the stochastic fluctuation remains. This leads to the calm zones of the intermittent acceleration. Since the highly non-Gaussian statistics of acceleration is fully accounted for by $A_q$, the residual fluctuation is expected to be strictly Gaussian. This expectation is fully supported by experimental data. As concerns the intermittent bursts, they are predicted to result from ``quantum jumps" between the probability peaks, when the particle velocity crosses the zero minima, involving a  divergence of the acceleration component $A_q$. There too, this process can be followed in detail in the experimental data.

(8) Finally, this analytical approach has been completed by numerical simulations of the Schr\"odinger/$A_q$ process. They have allowed us to recover the experimentally observed  autocorrelation functions of acceleration magnitude and the observed exponents of the structure functions.

These various effects and the data analysis methods by which they have been identified will be explained in more detail in specific works to come \cite{NottaleLehner}. 

We conclude by emphasizing that the new approach based on scale relativity methods that we have developed here is not statistical in its essence, but partly deterministic (although based on an underlying stochastic fluctuation). When a given large scale eddy is established for some time, its corresponding potential is determined and so is the solution of the Schr\"odinger equation including this potential. Then $P_v$ is known and finally the derived acceleration component $A_q(v)$. In practice, the complexity of the evolution of the velocity $v(t)$ leads to a pseudo-random nature of $A_q[v(t)]$: this explains why the intermittent acceleration in turbulent fluids has been considered up to now to be fully stochastic in nature. Another aspect of the present work is that, besides the new insights it brings on the nature of turbulence, it provides, if confirmed, a laboratory experimental validation of the theory of scale relativity (in its ``macroquantum" aspects).\\

{\bf Acknowledgements}: we are extremely grateful to Dr. Nicolas Mordant for kind communication of his experimental data on Lagrangian measurements in turbulence.



\end{document}